\documentclass[aps,prd,showpacs,twocolumn,10pt,superscriptaddress,floatfix,nofootinbib,longbibliography]{revtex4-1}   
\usepackage{amssymb,amsfonts,amsmath,bm,bbm}   
\usepackage{graphicx}
\usepackage{dcolumn}
\usepackage{bm}
\usepackage{CJK}
\usepackage{minted}  
\usepackage[
colorlinks=true,
linkcolor=blue,
breaklinks=true,
urlcolor=blue,
citecolor=blue]{hyperref}
\usepackage{bbold}
\usepackage{epstopdf}

\newcommand{\HNUST}{\affiliation{
Hunan Provincial Key Laboratory of Intelligent Sensors and Advanced Sensor Materials, \\ School of Physics and Electronics, Hunan University of Science and Technology, Xiangtan 411201, China}} 
      
\newcommand{\Squ}{\affiliation{
College of Information Engineering, Suqian University, Suqian 223800, China}}

\newcommand{\HNUSTchem}{\affiliation{
School of Chemistry and Chemical Engineering, Hunan University of Science and Technology, Xiangtan 411201, China}}

\begin{document}
\preprint{APS/123-QED}

\title{
Probing the nonstrange quark star equation of state with compact stars and gravitational waves
}

\author{Shu-Peng Wang}
\HNUST  

\author{Zhen-Yan Lu}\email
{luzhenyan@hnust.edu.cn}
\HNUST

\author{Zhi-Jun Ma}\email{mazhijun@hnust.edu.cn}\HNUST\HNUSTchem

\author{Rong-Yao Yang}\email{ryyang@seu.edu.cn}
\HNUST

\author{Jian-Feng Xu}\email
{xujf@squ.edu.cn}
\Squ

\author{Xiangyun Fu}\HNUST

\date{\today}

\begin{abstract}
A recent study shows that incorporating a new term into 
the thermodynamic potential density, as required by the thermodynamic consistency criterion, can effectively resolve the thermodynamic inconsistency problems of the conventional perturbative QCD model. This additional term 
plays a crucial role in resolving inconsistencies at relatively low densities and becomes negligible at extremely high densities. Within this revised 
perturbative QCD model,  
we find that if we require only that the energy per baryon of up-down ($ud$) quark matter exceeds 930 MeV so as not to contradict the standard nuclear physics, the maximum mass of an $ud$ quark star allowed by the revised perturbative QCD model can reach up to 2.17$M_{\odot}$. From this perspective, the observed 2.14$M_{\odot}$ pulsar PSR J0740+6620 may be an $ud$ quark star. However, if we further impose the constraint that the tidal deformability of a 1.4$M_{\odot}$ $ud$ quark star must be consistent with the GW170817 event, the maximum mass allowed by the revised perturbative QCD model would decrease to no more than 2.08$M_{\odot}$.
Consequently, our results suggest that the compact object with a mass of 2.50-2.67$M_{\odot}$, as observed in the GW190814 event, cannot be an $ud$ quark star, according to the revised perturbative QCD model.

\end{abstract}

\maketitle

\section{Introduction} \label{sec:INTRODUCTION}

The advent of multi-messenger astronomy, particularly the groundbreaking detections of gravitational-wave events like GW170817 and GW190814 by the LIGO-Virgo Collaboration~\cite{LIGOScientific:2017vwq,LIGOScientific:2020zkf}, has revolutionized our ability to probe the nature of dense matter. These observations, alongside precise mass measurements of massive pulsars such as PSR J0740+6620~\cite{Riley:2021pdl} and PSR J0952-0607~\cite{Romani:2022jhd}, have placed  unprecedentedly stringent constraints on the equation of state of ultra-dense objects~\cite{Ferreira:2021pni,Kanakis-Pegios:2020kzp,Yang:2019rxn,Zhu:2022ibs,Dexheimer:2020rlp,
Yang:2021bpe,Nunes:2020cuz,Annala:2021gom,Lim:2020zvx}. 
These advances underscore the critical importance of investigating the thermodynamic properties of dense matter and its equation of state~\cite{Oertel:2016bki,MUSES:2023hyz,ReinkePelicer:2025vuh}. A precise equation of state is essential for understanding the internal structure of compact objects~\cite{Oertel:2016bki}, interpreting gravitational wave signals~\cite{LIGOScientific:2018cki}, and mapping the QCD phase diagram~\cite{Lu:2025qyf,Aarts:2023vsf}.

Astronomical observations place stringent constraints on the parameters of effective field theories and phenomenological models~\cite{Roupas:2020nua,Yang:2023haz,Yuan:2024hge,Roupas:2020nua,Tews:2018iwm,Li:2021crp,Oikonomou:2023otn,Pi:2022pjs,Tangphati:2024atj}. 
This is exemplified by a recent study that rigorously tested over 500 relativistic mean-field theories against a comprehensive dataset of stellar structure constraints~\cite{Brodie:2023pjw}. This dataset includes the mass measurement of the massive pulsar J0740+6620, 
multimessenger constraints combining NICER and XMM-Newton observations~\cite{Pang:2021jta}, tidal deformability from the gravitational-wave events GW170817 and GW190425, detailed modeling of the kilonova AT2017gfo and the gamma-ray burst GRB170817A, and the remarkable mass-radius measurement of the compact object HESS J1731-347~\cite{Doroshenko:2022nwp}. The formidable challenge posed by these multi-faceted constraints is such that none of the sampled conventional nuclear models can simultaneously satisfy all observational requirements at the 68\% credibility level. Consequently, this profound challenge has spurred investigations into alternative compact star models, namely those with a quark-matter core~\cite{Pal:2025skz,Ayriyan:2024zfw,Pal:2024afl,Mariani:2022xek,Ferreira:2021osk,Ranea-Sandoval:2019miz,Li:2021sxb,Albino:2024ymc,Blaschke:2022egm} and those composed entirely of quark matter~\cite{Tangphati:2024ycu,Xu:2022squ,Banerjee:2025ufb,Tangphati:2022arm,Carvalho:2022kxq}.

At low energies, due to the notorious sign problem prohibiting lattice QCD calculations at finite baryon densities, effective field theories and QCD phenomenological models serve as essential tools for probing dense matter~\cite{Drischler:2021kxf,Pretel:2025roz,Pretel:2023nlr,Tangphati:2023efy,Tangphati:2022acb,Yuan:2022dxb,Xia:2022dvw,Chu:2025tsx,Liu:2023ocg} and QCD phase transitions. 
When the density becomes sufficiently high, the breakdown of color confinement leads to quark deconfinement, and nuclear matter is expected to transform into quark matter that can realize a rich variety of phases, including color-superconducting states~\cite{Alford:2007xm,Fukushima:2010bq,gartlein2025Colorsuperconducting,Ivanytskyi:2022oxv}.
For phenomenological models, ensuring thermodynamic self-consistency is a critical issue that must be carefully addressed~\cite{Wen:2005uf,Chu:2012rd,Lu:2016fki,Alba:2014lda,Pal:2023dlv,Xu:2015wya}. One of the most commonly used methods for describing quark matter is perturbative QCD, which provides a robust framework for high-density conditions~\cite{Carignano:2016rvs,Graf:2015pyl}. However, extrapolating perturbative QCD results to the lower densities relevant to astrophysical environments introduces severe thermodynamic inconsistencies. 
This issue arises from the artificial extrapolation of a perturbative calculation, which does not exhibit thermodynamic problems within its valid perturbative regime, to a non-perturbative system without considering non-perturbative effects. Reliable models must ensure that the pressure vanishes at the minimum energy per baryon, a condition that conventional perturbative QCD model fails to meet~\cite{Peng:2005xp,Xu:2015wya}.

By incorporating a new term determined by thermodynamic consistency requirement into the thermodynamic potential density, it has been shown that this approach can effectively resolve the thermodynamic inconsistency problems of the conventional perturbative QCD model~\cite{Xu:2014zea}. 
This additional term plays a crucial role at lower densities while being negligible at high densities. This approach has been shown to yield a more reasonable density dependence for the sound velocity. In this work, we adopt this revised perturbative QCD model to constrain the equation of state and its parameter space for $ud$ quark matter. The constraint is achieved by combining the tidal deformability data from the GW170817 event and the mass-radius observations of massive compact stars. Building on the constrained equation of state, we further investigate the maximum mass allowed by the model and explore the possibility that the 2.50-2.67 $M_\odot$ compact object in the GW190814 event could be an $ud$ quark star. As we will see, under the condition that the energy per baryon of $ud$ quark matter is greater than 930 MeV, the maximum mass of a compact star supported by the revised perturbative QCD model is approximately 2.17 $M_\odot$. When the tidal deformability observation from the GW170817 event $\tilde{\Lambda}_{1.4}=190_{-120}^{+390}$ is introduced as an additional constraint, this maximum allowable mass is further reduced to 2.07 $M_\odot$. 

The paper is organized as follows. In Sec.~\ref{sec:pQCD}, we describe the perturbative thermodynamics of cold QCD with the low-density correction.  
In Sec.~\ref{sec:numericalresult}, we present our results, including thermodynamic properties of $ud$ quark matter and the mass-radius relations and tidal deformabilities of $ud$ quark stars, and compare them with observational constraints.
Finally, a short summary is given in Sec.~\ref{sec:CONCLUSION}.

\section{Thermodynamic self-consistent perturbative QCD model at finite chemical potential}\label{sec:pQCD}

The thermodynamic potential density of the up and down quark system, taking into account first-order corrections to the coupling constant, is given by
\begin{align} \label{eq:Omegapt}
\Omega_{\mathrm{pt}}=-\frac{1}{4 \pi^2}\left(\mu_{u}^4+\mu_{d}^4\right)\left(1-
2\alpha
\right),
\end{align}
where $\mu_{u}$ and $\mu_{d}$ represent the chemical potentials for the up and down quarks, respectively, and $\alpha=\alpha_s/\pi$
with 
the explicit expression for $\alpha$ is given by
\begin{align} \label{eq:alpha}
\alpha=\frac{1}{\beta_{0}\ln(\Lambda/\Lambda_{\mathrm{QCD}})},
\end{align}
where 
$\beta_0=11/2-N_f/3$, and the QCD scale parameter $\Lambda_{\mathrm{QCD}}$ is taken to be $\Lambda_{\mathrm{QCD}}=$ 147 MeV~\cite{Xu:2014zea}. 
It is important to emphasize that the relationship between the renormalization subtraction point and the quark chemical potential cannot be chosen arbitrarily, as this may lead to  thermodynamic inconsistencies. For instance, some previous studies have used the renormalization subtraction point as an average chemical potential, $\bar{\mu}$, namely~\cite{Fraga:2001id,Fraga:2001xc,Fraga:2004gz}
\begin{align}
    \Lambda/\bar{\mu}=1,~2,~3,
\end{align}
and in Ref.~\cite{Freedman:1976ub}, it is assumed to be
\begin{align}
   \Lambda=\sqrt{(\mu_u^2+\mu_d^2)/2}. 
\end{align}
However, conventional perturbative QCD models neglect higher-order terms in $\alpha$, which can lead to thermodynamic inconsistency~\cite{Xu:2015wya}. Therefore, to perform concrete calculations, it is necessary to determine the relationship between the renormalization subtraction point and the chemical potentials.
As mentioned in the Introduction, we need to include a new contribution $\Omega^{\prime}$ into the thermodynamic potential density of the system, and ensure that the integral of $\Omega^{\prime}$ is path-independent. Based on this, the renormalization subtraction point $\Lambda$, determined by satisfying the fundamental differential equation of thermodynamics, is given by~\cite{Xu:2014zea} 
\begin{align}
\Lambda=
C
\left(\frac{\mu_{u}^{4}+\mu_{d}^{4}}{2}\right)^{1/4} ,
\end{align}
where $C$ is a dimensionless model parameter and the number 2 in the denominator represents the number of quark flavors. 
Using this equation, it has been demonstrated that the minimum energy of the quark matter system coincides with the zero point of the pressure, thus satisfying the thermodynamic requirement for self-consistency~\cite{Xu:2014zea}.
It is also worth mentioning that, in the quasiparticle model, the relationship between the renormalization subtraction point and the quark chemical potentials cannot be chosen arbitrarily; instead, it must be determined by the thermodynamic consistency requirements of the model~\cite{Ma:2023stj,Lu:2016fki}.


As stated above, the conventional perturbative QCD model suffers from the thermodynamic inconsistency problem, and the problem can be addressed by introducing an additional term, $\Omega^\prime$, to the thermodynamic potential density given in Eq.~(\ref{eq:Omegapt}), which is determined by meeting the fundamental differential equation of thermodynamics, and given by 
\begin{equation} \label{eq:OmegaPrime}
\Omega^{\prime}=\int_{(\mu_{u0},\mu_{d0})}^{(\mu_u,\mu_d)}
\left[\mathcal{F}(\mu_{u},\mu_{d})\mathrm{d}\mu_{u}+\mathcal{Q}(\mu_{u},\mu_{d})\mathrm{d}\mu_{d}\right]+B_{0},
\end{equation}
where $\mu_{\text{u}0}$ and $\mu_{\text{d}0}$ denote the starting points for the integral, which are set to $\mu_{\text{u}0}=\mu_{\text{d}0}=300$ MeV in our calculation, while their shift effect is absorbed into the constant vacuum pressure $B_0$
~\cite{Farhi:1984qu,Lopes:2020btp,Pal:2023dlv,Pal:2023quk}, and the auxiliary physical quantities 
$\mathcal{F}(\mu_{u},\mu_{d})$ and $\mathcal{Q}(\mu_{u},\mu_{d})$ are given by
\begin{align}
\mathcal{F}(\mu_{u},\mu_{d})
=
\frac{\mu_{u}^{3}}{2\beta_{0}\pi^{2}}\ln^{-2}\left(\frac{C}{\Lambda_{\mathrm{QCD}}}\sqrt[4]{\frac{\mu_{u}^{4}+\mu_{d}^{4}}{2}}\right) ,
\end{align}
and
\begin{align}
\mathcal{Q}(\mu_{u},\mu_{d})
=
\frac{\mu_{d}^{3}}{2\beta_{0}\pi^{2}}\ln^{-2}\left(\frac{C}{\Lambda_{\mathrm{QCD}}}\sqrt[4]{\frac{\mu_{u}^{4}+\mu_{d}^{4}}{2}}\right).
\end{align}
Then, the total thermodynamic potential density of $ud$ quark matter is obtained as $\Omega=\Omega_\text{pt}+\Omega^{\prime}$, namely 
\begin{eqnarray} \label{eq:Omegafull}
  \Omega&=&-\frac{1}{4\pi^2}\Big(\mu_u^4+\mu_d^4\Big)(1-2\alpha)-\frac{\mu_e^4}{12\pi^2}+B_0 
\nonumber\\
&&+\int_{(\mu_{u0},\mu_{d0})}^{{(\mu_u,\mu_d)}}\Bigg[
\frac{\mu_{u}^{3}}{2\beta_{0}\pi^{2}}\ln^{-2}\left(\frac{C}{\Lambda_{\mathrm{QCD}}}\sqrt[4]{\frac{\mu_{u}^{4}+\mu_{d}^{4}}{2}}\right)\text{d}\mu_u\nonumber\\
&&+
\frac{\mu_{d}^{3}}{{2\beta_{0}\pi^{2}}}\ln^{-2}\left(\frac{C}{\Lambda_{\mathrm{QCD}}}\sqrt[4]{\frac{\mu_{u}^{4}+\mu_{d}^{4}}{2}}\right)\text{d}\mu_d
\Bigg].~~~
\end{eqnarray}
We emphasize that, due to the complexity and length of the integrand in Eq.~(\ref{eq:OmegaPrime}), only numerical methods are feasible. Therefore, the integral form of the additional term $\Omega^\prime$ is retained in Eq.~(\ref{eq:Omegafull}). 
The number density of quark flavor $i$ and electrons can be obtained by taking the first derivative of the corresponding thermodynamic potential with respect to the chemical potential, which gives
\begin{align}
   n_{q}=\frac{1}{\pi^{2}}\mu_{q}^{3}(1-2\alpha), 
~~~n_e=\frac{1}{3\pi^2}\mu_e^3,
\end{align}
respectively. With Eq.~(\ref{eq:Omegafull}) at hand, the energy density and the pressure of the system can then be derived as
\begin{align}\label{eq:E_P}
E=\Omega+\sum_{i}\mu_{i}n_{i},
~~~~~~P=-\Omega.
\end{align}

\section{Numerical results and discussions}\label{sec:numericalresult}

Considering a system composed of up and down quarks and electrons 
in $\beta$ equilibrium, which is achieved through the weak reactions $d\leftrightarrow u+e+\bar{\nu}_e$, one easily obtains the following equality for the chemical potential equilibrium condition
\begin{align}\label{eq:muUDe}
\mu_d=\mu_u+\mu_{e}.
\end{align}
Moreover, neutrinos are assumed to freely escape and enter the system, so their chemical potential is taken to be zero. 
The baryon density $n_b$ is defined as
\begin{align}\label{eq:nB}
n_b=\frac{1}{3}(n_u+n_d).
\end{align}
The charge neutrality condition is also imposed on the system, that is,
\begin{align}\label{eq:nQ0}
\frac{2}{3}n_u-\frac{1}{3}n_d-n_{\mathrm{e}}=0.
\end{align}

For a given baryon density $n_b$, the chemical potentials $\mu_i$ can be obtained by solving the sets of Eqs.~(\ref{eq:muUDe})-(\ref{eq:nQ0}). Subsequently, we can perform numerical calculations on Eq.~(\ref{eq:Omegafull}) to determine the energy density and pressure of the system. This allows us to further characterize the thermodynamic properties of the system and investigate the properties of quark matter as well as the structure of quark stars.

\subsection{Stability window for $ud$ quark matter}

\begin{figure}[h]  \includegraphics[width=0.48\textwidth]{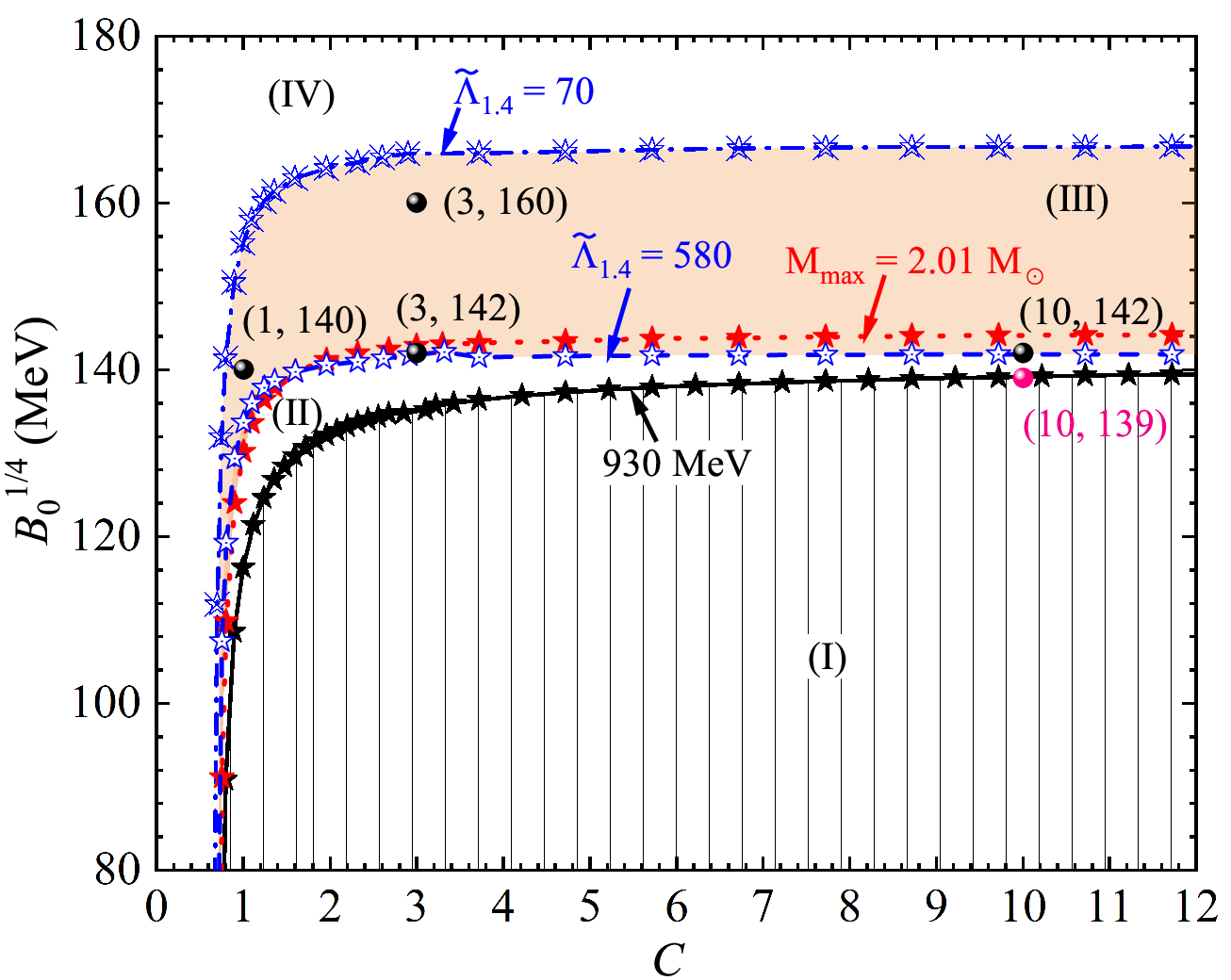}\\
\caption{Stability window for $ud$ quark matter in the $C-B_0^{1/4}$ plane. 
The bottom-right black shaded area is forbidden, where the energy per baryon of $ud$ quark matter is less than 930 MeV. Meanwhile, the orange shaded region between the blue dashed and blue dash-dotted curves with star symbols corresponds to the $ud$ quark matter stable region with an equation of state that can support an $ud$ quark star with the tidal deformability of a $1.4~M_{\odot}$ quark star in the range $70\leq  \tilde{\Lambda}_{1.4}\leq 580$. The selected typical model parameters are indicated
with black solid dots.
}\label{fig:C0B0}
\end{figure}

In order not to contradict the standard nuclear physics, we note that the energy per baryon of $ud$ quark matter should not be less than 930 MeV, namely
\begin{align}
    \epsilon_{ud}=\frac{E}{n_b}\bigg|_{ud} \geq 930~\text{MeV} .
\end{align}
In this case, the stable condition of $ud$ quark matter, along with astronomical observations of the tidal deformability and massive masses of compact stars, impose strict constraints on the parameter space of the revised perturbative QCD model. In Fig.~\ref{fig:C0B0}, we show the stability window of $ud$ quark matter in the $C-B_0^{1/4}$ plane, where the region above the black solid curve with full black stars corresponds to the allowed region for the $ud$ quark matter, while the bottom-right shaded area with an energy per baryon of less than 930 MeV is forbidden. In addition, the measured gravitational mass of PSR J0740$+$6620 with a lower limit of 2.01$M_{\odot}$ is also plotted 
in the $C-B_0^{1/4}$ plane, represented by the red dotted curve with star symbols. We note that the maximum mass of quark stars can only exceed 2.01$M_{\odot}$ when the parameter space lies below the red dashed curve with star symbols. Furthermore, for  a given value of $C$, the smaller the value of $B_0$, the larger the maximum mass of the corresponding compact star.

For the GW170817 event, the observed tidal deformability is $\tilde{\Lambda}_{1.4}=190_{-120}^{+390}$, with $\tilde{\Lambda}_{1.4}$ denoting the dimensionless tidal deformability of a $1.4~M_{\odot}$ star. This value is illustrated by the orange shaded region labeled as Region III, which is delineated by the blue dash-dotted and blue dashed curves with star symbols. As shown in Fig.~\ref{fig:C0B0}, the red dotted curve with star symbols falls within the orange shaded area. This suggests that, within the revised perturbative QCD model, $ud$ quark stars can 
explain both the gravitational mass of PSR J0740+6620 and the tidal deformability observed in the GW170817 event.

To facilitate the following discussion, in addition to the orange region designated as Region III, we further label the bottom-right black shaded area as Region I, the white area between Region I and Region III as Region II, and the white area top-left of Region III as Region IV. According to this classification, Region I is forbidden because the energy per baryon of the $ud$ matter in this region is less than 930 MeV. Regions II, III, and IV are permitted; 
however, only the equation of state corresponding to Region III
yields a tidal deformability that is consistent with the observational data from the GW170817 event.

It is worth noting that the present study does not perform a Bayesian exploration of the parameter space for 
$C$ and $B_0$ shown in Fig.~\ref{fig:C0B0}, which would enable a more quantitative assessment of their uncertainties and correlations and, in turn, strengthen the statistical robustness of the observational constraints. 
Nevertheless, Bayesian inference techniques have been extensively applied in recent studies of compact stars and the dense-matter equation of state, e.g. Refs.~\cite{Hippert:2023bel,Marquez:2024bzj,Malik:2022jqc,Drischler:2024ebw,Semposki:2024vnp,Drischler:2020hwi}.
These methods have proven particularly powerful for constraining model parameters and quantifying uncertainties in multi-messenger analyses, offering valuable methodological guidance for future extensions of the present work.

\subsection{Properties of $ud$ quark matter}

\begin{figure}[h]
  \includegraphics[width=0.48\textwidth]{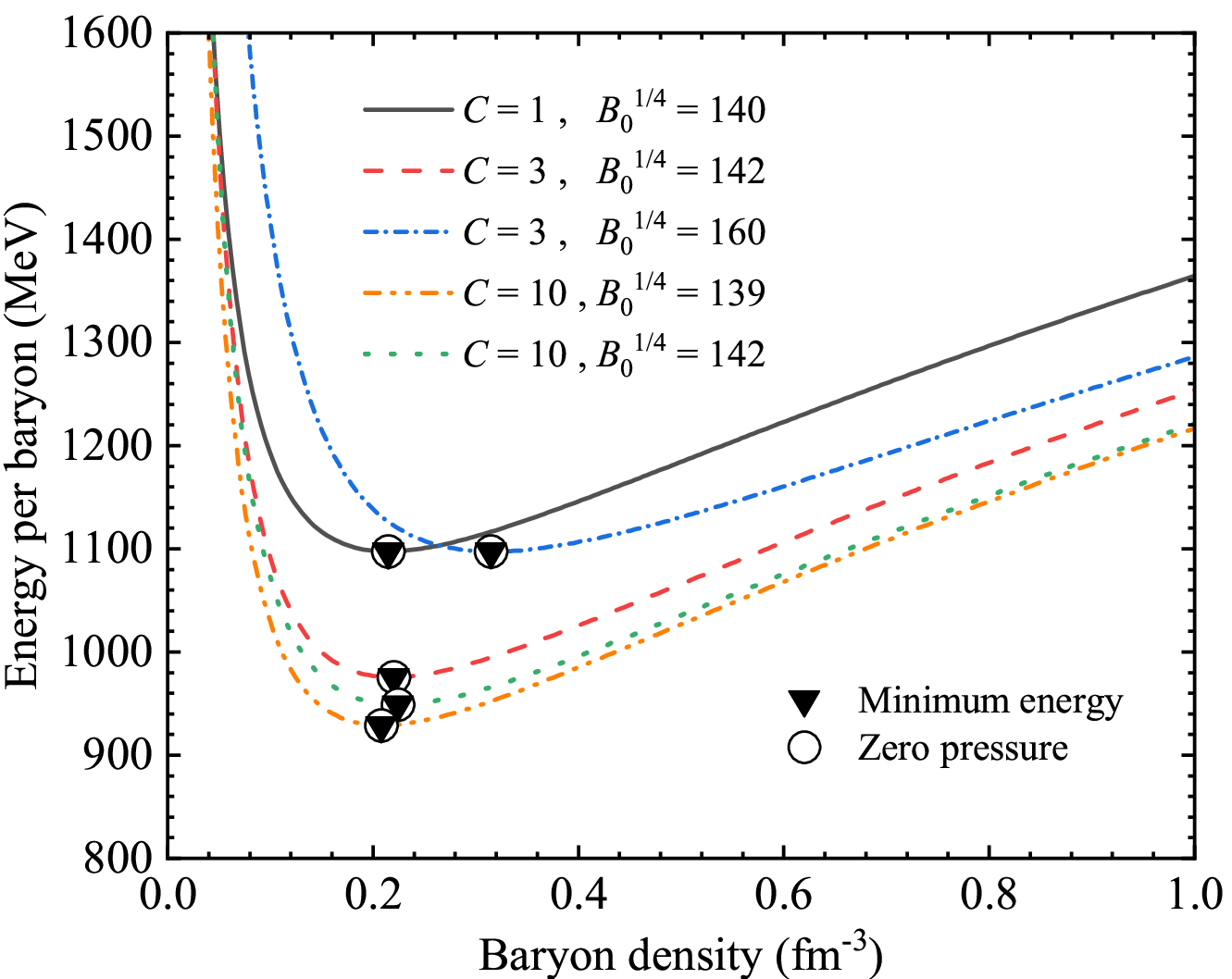}\\
  \caption{
Density behavior of the energy per baryon for different $C$ and $B_0$ values. The minimum energy and zero pressure are marked with inverted triangles and open circles, respectively.
}\label{fig:NbEnergy}
\end{figure}

To investigate how the parameters $C$ and $B_0$ influence the properties of $ud$ quark matter, we select five representative parameter sets, indicated by the solid dots in Fig.~\ref{fig:C0B0}, given by $(C, B_0^{1/4} / \text{MeV}) =$ (1, 140), (3, 142), (3, 160),  (10, 139), and (10, 142). 
In Fig.~\ref{fig:NbEnergy}, we plot the energy per baryon as a function of baryon density for the selected typical parameter sets. 
Within the revised perturbative QCD model, it can be seen that the point at which the pressure vanishes coincides precisely with the minimum energy per baryon. This feature ensures that quark matter remains stable at the appropriate density and that the model satisfies the requirements of thermodynamic consistency. 
By comparing the red dashed curve and the blue dash-dotted curve, both of which correspond to $C=3$, we observe that the blue dash-dotted curve, which has a larger $B_0$ value, lies generally above the red dashed curve with a smaller $B_0$. This  indicates that, at a given baryon density, a larger constant vacuum pressure $B_0$ leads to a higher energy per baryon for quark matter.  
On the contrary, as shown by the red dashed and green dotted curves, when $B_0^{1/4}$ is fixed at 142 MeV and $C$ is set to 3 and 10, respectively, the red dashed curve corresponding to the smaller $C$ value generally lies above the green dotted curve with a larger $C$.
This is because, according to Eq.~(\ref{eq:alpha}), a larger $C$ results in a smaller running coupling constant, which in turn reduces the thermodynamic potential density and energy density of the system.

\begin{figure}[h]
  \includegraphics[width=0.48\textwidth]{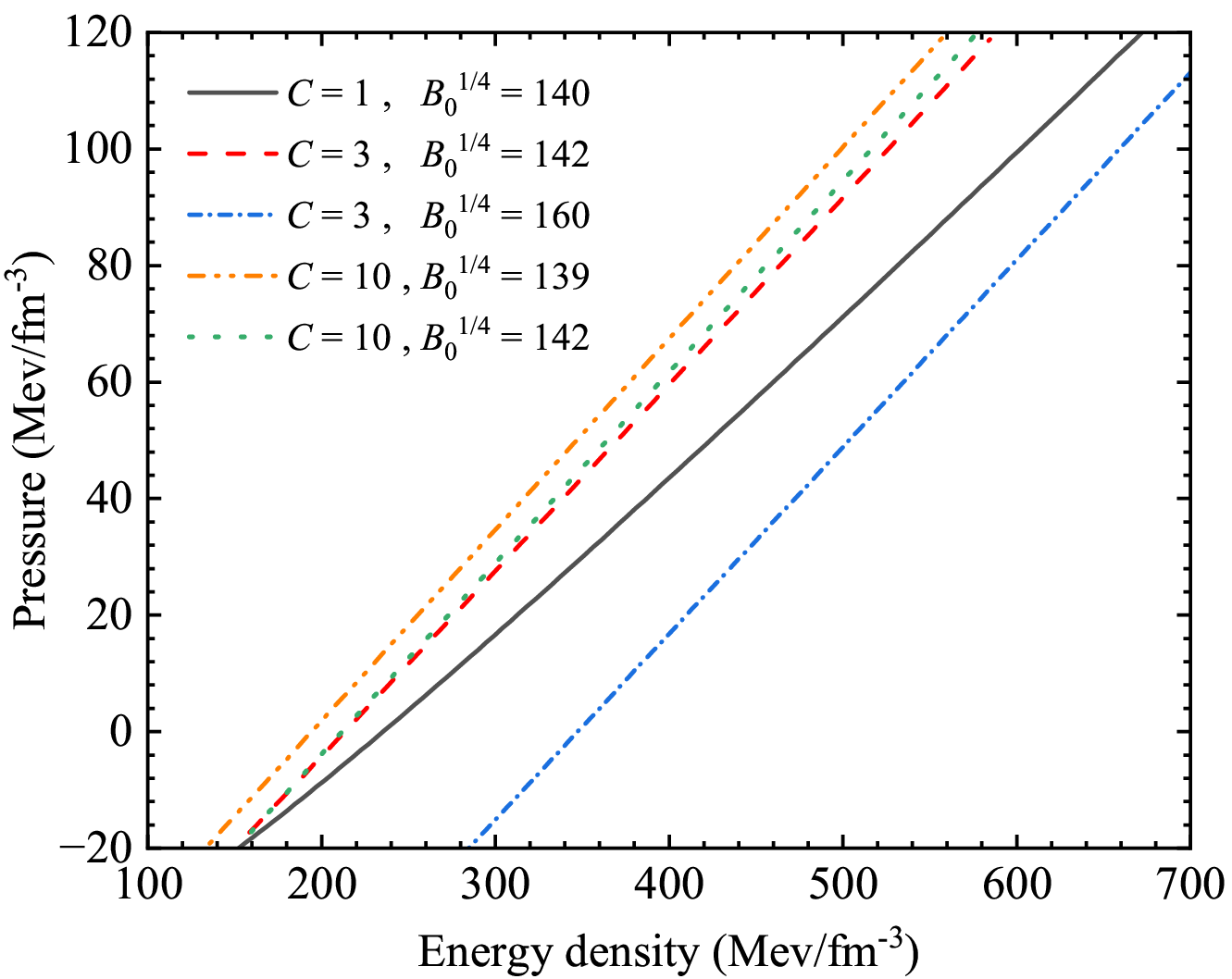}\\
  \caption{Equation of state of $ud$ quark matter for the same parameter sets as Fig.~\ref{fig:NbEnergy}.  
}\label{fig:EnergyPressure}
\end{figure}

The equation of state of dense matter is essential for calculating the mass-radius relationship and tidal deformability of compact stars. In Fig.~\ref{fig:EnergyPressure}, we show the pressure of $ud$ quark matter as a function of the energy density for the same parameter sets as Fig.~\ref{fig:NbEnergy}. Although all five curves show an approximately linear increase with energy density, it can be observed that the black curve, which corresponds to the parameters $C = 1$ and $B_0^{1/4} = 140$ MeV, exhibits a significantly slower growth rate. This suggests that the equation of state represented by this curve is relatively soft. 
For the red dashed curve and the green dotted curve, which share the same constant vacuum pressure, $B_0^{1/4} = 142$ MeV, the pressure growth rate increases as $C$ increases, indicating that the equation of state   becomes stiffer with increasing $C$. Furthermore, comparing the blue dash-dotted curve and the red dash-dotted curve, both of which have $C=3$, reveals that the red dashed curve, which corresponds to a smaller $B_0$, represents a stiffer equation of state.


\begin{figure}[h]
\includegraphics[width=0.48\textwidth]{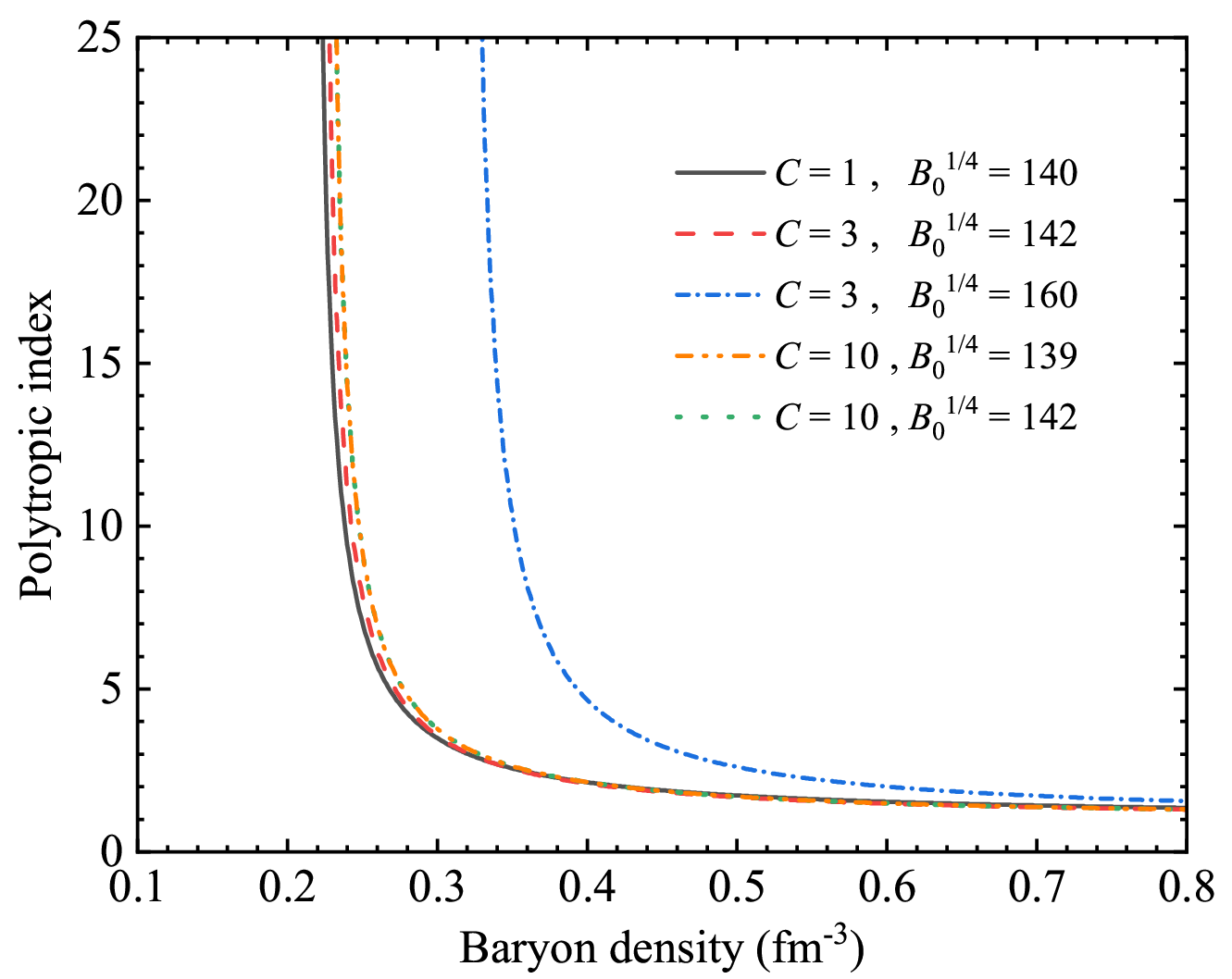}\\
\caption{The polytropic index as a function of the baryon density at zero temperature for the same parameter sets as Fig.~\ref{fig:NbEnergy}.
}\label{Fig:gamma-nb}
\end{figure}

The polytropic index, $\gamma$, is used to determine the inner composition of stellar matter. 
It is mathematically defined as
\begin{equation}
\gamma=\frac{\partial \ln P}{\partial \ln E},
\end{equation}
and uses the equation of state as its input. For matter exhibiting exact conformal symmetry and lacking intrinsic scales,  $\gamma=1$, irrespective of the coupling strength. This symmetry implies that the matter is independent of any dimensionful parameter, resulting in the energy density becoming proportional to the pressure, which consequently leads to $\gamma=1$.

In Fig.~\ref{Fig:gamma-nb}, we show the polytropic index as a function of baryon density at zero temperature within the revised perturbative QCD model for different $C$ and $B_0$. 
For the $ud$ quark matter, we observe that the polytropic index decreases monotonically with increasing density in all cases. Notably, within a very narrow density range, the polytropic index undergoes a sharp decline before stabilizing and then continuing to decrease at higher densities. 
More specifically, except for the blue dash-dotted curve, which corresponds to the largest values of $B_0$, the other four curves exhibit very similar behavior within the considered density range. Notably, the sharp drop in the polytropic index for the blue dash-dotted curve occurs at a higher density. 
Comparing the points at which the blue curve and the other curves undergo a sharp decline reveals that the location of the abrupt drop in the polytropic index is more sensitive to the value of $B_0$ than to the $C$ parameter. 
As can also be observed in Fig.~\ref{Fig:gamma-nb}, within the revised perturbative QCD model, the polytropic index $\gamma$ decreases to a lower limit of $\gamma = 1.2$ for all cases as the baryon density $n_b$ increases to 0.8 fm$^{-3}$. 
In addition, we also verified that as the baryon density $n_b$ continues to increase, the polytropic index gradually decreases and approaches 1. 
This finding is consistent with the results reported in Ref.~\cite{Annala:2019puf}, where $\gamma < 1.75$ was found for quark matter at high baryon densities.

\subsection{$ud$ quark stars}

\begin{figure}[h]
  \includegraphics[width=0.48\textwidth]{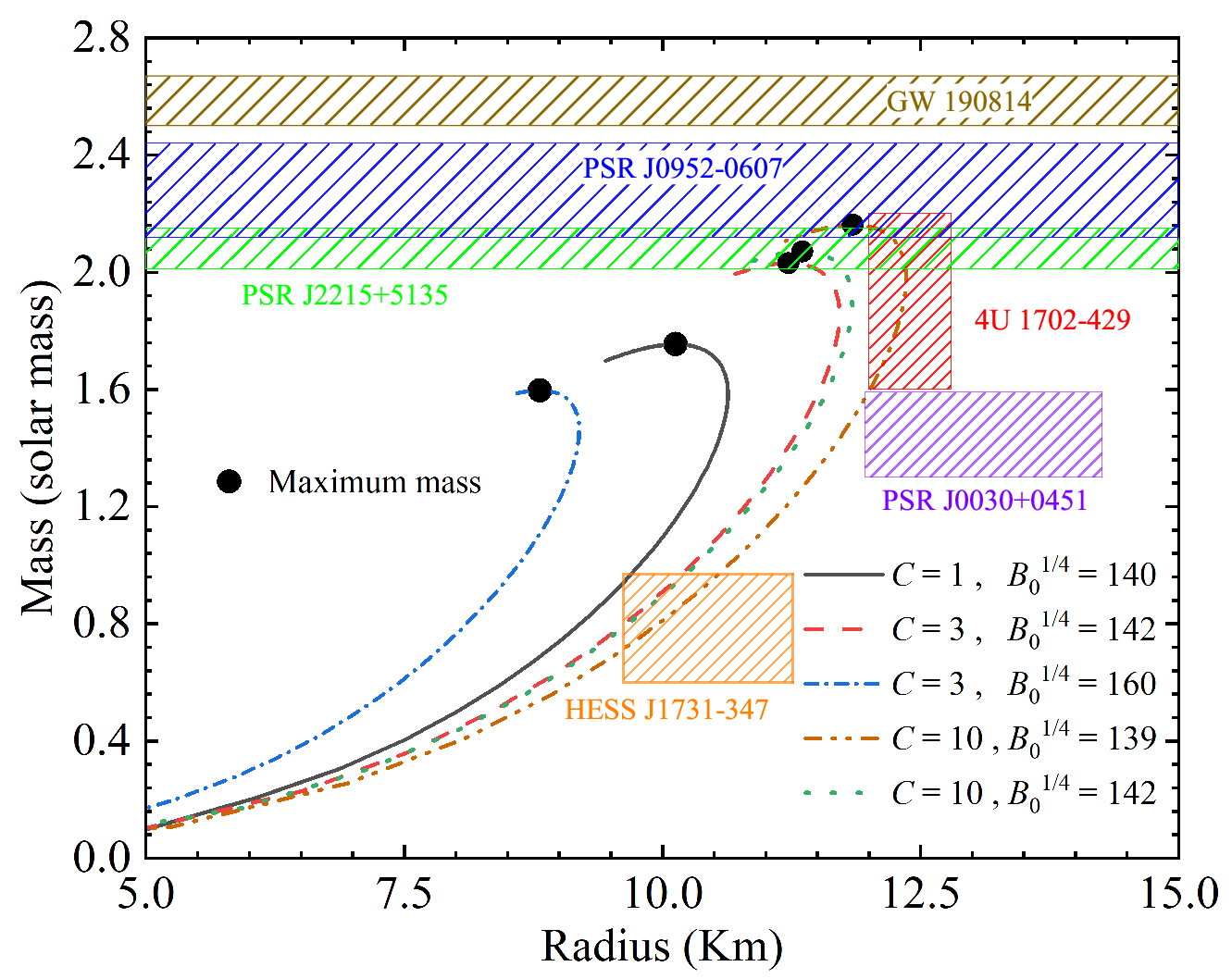}\\
  \caption{Mass-radius relations of quark stars for the same parameter sets as Fig.~\ref{fig:NbEnergy}. The maximum masses of the $ud$ quark stars are indicated by black solid circles. 
}\label{fig:MassRadius}
\end{figure}


The mass-radius relationship of a quark star is a key aspect in understanding the structure and properties of these hypothetical compact objects, which are proposed to be composed of quark matter rather than the neutron-rich matter found in neutron stars. In this work, we model compact stellar objects as systems consisting entirely of $ud$ quark matter and electrons, hereafter referred to as $ud$ quark stars. 
The equilibrium structure of a static spherically symmetric quark star is determined by the Tolman-Oppenheimer-Volkoff (TOV) equation~\cite{Oppenheimer:1939ne}
\begin{eqnarray}
   \frac{\mathrm{d} P(r)}{\mathrm{d} r}=-\frac{G m E}{r^2} \frac{(1+P / E)\left(1+4 \pi r^3 P / m\right)}{1-2 G m / r},
\end{eqnarray}
and the subsidiary condition
\begin{eqnarray}
  \frac{\mathrm{d} m(r)}{\mathrm{d} r}=4 \pi r^2 E.
\end{eqnarray}
Here, $G=6.7 \times 10^{-45}~ \mathrm{MeV}^{-2}$ denotes the gravitational constant, $r$ represents the radial coordinate of the quark star, and $m$ corresponds to the gravitational mass enclosed up to the radius $r$.

By numerically solving the coupled TOV equations with the equation of state presented in Fig.~\ref{fig:EnergyPressure} as input, we obtain the mass-radius relationship for quark stars. In
Fig.~\ref{fig:MassRadius}, we show the mass-radius curves for $ud$ quark stars computed using different parameter sets. 
For comparison, we also included recent observational data for the gravitational mass, radius, and tidal deformability of compact stars from the literature. 
The brown shaded region marks the mass measurement of GW190814's secondary component, which at $2.59^{+0.08}_{-0.09}~M_{\odot}$ represents one of the heaviest compact stars ever detected~\cite{LIGOScientific:2020zkf}. The blue and green regions correspond to the millisecond pulsars PSR J0952-0607 and PSR J2215+5135, respectively, with precisely measured masses of $M=2.35 \pm 0.17~M_{\odot}$~\cite{Romani:2022jhd} and $M=2.27_{-0.15}^{+0.17}~M_{\odot}$~\cite{Linares:2018ppq}. The red shaded region shows constraints derived from X-ray observations of 4U 1702-429, yielding $M=1.9\pm0.3~M_{\odot}$ and $R=12.4\pm 0.4$ km~\cite{Nattila:2017wtj}. Meanwhile, the purple region displays the NICER results for PSR J0030+0451, representing $M=1.44_{-0.14}^{+0.15}~M_{\odot}$ and $R=13.02_{-1.06}^{+1.24}$ km~\cite{Miller:2019cac}. Finally, the orange region depicts the unusually compact star embedded in the supernova remnant HESS J1731-347, with $M=0.77_{-0.17}^{+0.20}~M_{\odot}$ and $R=10.4_{-0.78}^{+0.86}$ km~\cite{Doroshenko:2022nwp}. Together, these observations span nearly an order of magnitude in mass and significant variation in radius, providing critical benchmarks for testing theoretical equations of state and understanding the nature of matter at high densities.

As illustrated in the figure, within the revised perturbative QCD model, the maximum masses corresponding to the blue dash-dotted and black solid lines are significantly below 2$M_{\odot}$, failing to satisfy the observational constraints on masses and radii from six compact stars. While the parameters of these two curves lie within the orange region that satisfies the GW170817 tidal deformation constraint, they cannot conform to all of the observational data from compact stars simultaneously.
In contrast, the red dashed and green dotted lines corresponding to $C=3$, $B_0^{1/4}=142$ MeV and $C=10$, $B_0^{1/4}=142$ MeV respectively, can support maximum masses of 2.03$-$2.07$M_{\odot}$ and pass through the mass-radius observational range of HESS J1731-347. However, these two curves still fail to cover the observational data for 4U 1702-429 and PSR J0030+0451.
Notably, the orange dotted-dashed line with $C=10$ and  $B_0^{1/4}=139$ MeV, does not satisfy the GW170817 tidal deformation constraint, but completely traverses the orange, purple, and red shaded regions in Fig.~\ref{fig:MassRadius}.
So, to simultaneously satisfy the observational data from the mentioned compact stars, the model parameters should be selected from Region II in Fig.~\ref{fig:C0B0} with $C>1$. Within this parameter range, the model can yield a maximum mass of 2$-$2.1$M_{\odot}$, which is consistent with the observational data for HESS J1731-347, PSR J0030+0451, and 4U 1702-429. However, it should be noted that no combination of $C$ and $B_0$ values can produce a stellar maximum mass exceeding 2.2$M_{\odot}$ in the parameter range with the constraint $\epsilon_{ud}\geq 930$ MeV. This implies that in the revised perturbative QCD model, the reported compact star with a mass of $2.59^{+0.08}_{-0.09}~M_{\odot}$ from the GW190814 event cannot be an $ud$ quark star. This conclusion contrasts sharply with those recently recent conclusions drawn from a confining quark matter model~\cite{Cao:2020zxi}.

Beyond the mass-radius relation, the recent detection of gravitational waves from event GW170817~\cite{LIGOScientific:2018cki} has provided stringent experimental constraints on the equation of state for compact stars through measured tidal deformability~\cite{Drischler:2020fvz,Contrera:2022tqh,Yang:2021sqg}. In recent years, significant efforts have been devoted to constraining the properties of quark star matter, particularly based on the improved estimate of tidal deformability $\tilde{\Lambda}_{1.4}=190^{+390}_{-120}$ from GW170817, where $\tilde{\Lambda}_{1.4}$ denotes the dimensionless tidal deformability of a $1.4~M_{\odot}$ compact star.
The response of quark stars to gravitational fields is characterized by the tidal Love number $k_2$, which depends on the stellar structure and consequently on the mass and equation of state of quark matter. In General Relativity, the dimensionless tidal deformability $\tilde{\Lambda}$ is related to the $l=2$ tidal Love number $k_2$ as follows~\cite{Hinderer:2007mb,LIGOScientific:2018cki}
\begin{eqnarray}
\tilde{\Lambda}=
\frac{2}{3\beta^5} k_2,
\end{eqnarray}
where $\beta=GM/R$ and $R$ are the compactness parameter and radius of the star, respectively. 
The Love number $k_2$ can be expressed as~\cite{Flanagan:2007ix}  
\begin{eqnarray}
k_2&=& \frac{8 \beta^5}{5}(1-2 \beta)^2[2+2 \beta(y_R-1)-y_R] \nonumber\\
&& \times\{
4 \beta^3\left[13-11 y_R+\beta(3 y_R-2)+2 \beta^2(1+y_R)\right] \nonumber\\
&& +3(1-2 \beta)^2[2-y_R+2 \beta(y_R-1)] \ln (1-2 \beta) \nonumber\\
&&+2 \beta[6-3 y_R+3 \beta(5 y_R-8)] 
\},~~~~~
\end{eqnarray}
where 
$y_R\equiv y(R)$, and $y(r)$ is determined by solving the following differential equation
\begin{eqnarray}
r \frac{d y(r)}{d r}+y^2(r)+y(r) F(r)+r^2 Q(r)=0.
\end{eqnarray}
Here, $F(r)$ and $Q(r)$ are, respectively, defined as
\begin{eqnarray}
F(r) \equiv \frac{1-4 \pi r^2[E(r)-P(r)] G}{f(r)}
\end{eqnarray}
and
\begin{eqnarray}
Q(r) &\equiv & \frac{4 \pi}{f(r)}\bigg[5 E(r) G+9 P(r) G+\frac{E(r)+P(r)}{V_s^2(r)} G \nonumber\\
&& -\frac{6}{4 \pi r^2}\bigg]-4\left[\frac{m(r)+4 \pi r^3 P(r)}{r^2 f(r)} G\right]^2,
\end{eqnarray}
where $V_s^2$ represents the squared sound velocity of quark matter, and $f(r)$ takes the form $f(r)=1-2 Gm(r) / r$.   

Our findings illustrate that the revised perturbative QCD model employed in this study retains the high-density predictions of standard perturbative QCD while mitigating its pathological behavior at low densities through the incorporation of thermodynamic corrections. This dual functionality allows the model to accommodate quark stars with masses up to $2 M_{\odot}$, in agreement with observations of massive pulsars such as PSR J0740+6620~\cite{Xu:2014zea}. For a $1.4~M_{\odot}$ quark star, the predicted tidal deformability $\tilde{\Lambda}_{1.4}$ ranges from 70 to 580, aligning with the constraints imposed by GW170817~\cite{LIGOScientific:2018cki}. These results offer valuable theoretical insights into the internal structure of compact stars and their gravitational wave signatures, thereby bridging the gap between quark matter theory and multi-messenger astronomy.

\begin{figure}[h]
  \includegraphics[width=0.48\textwidth]{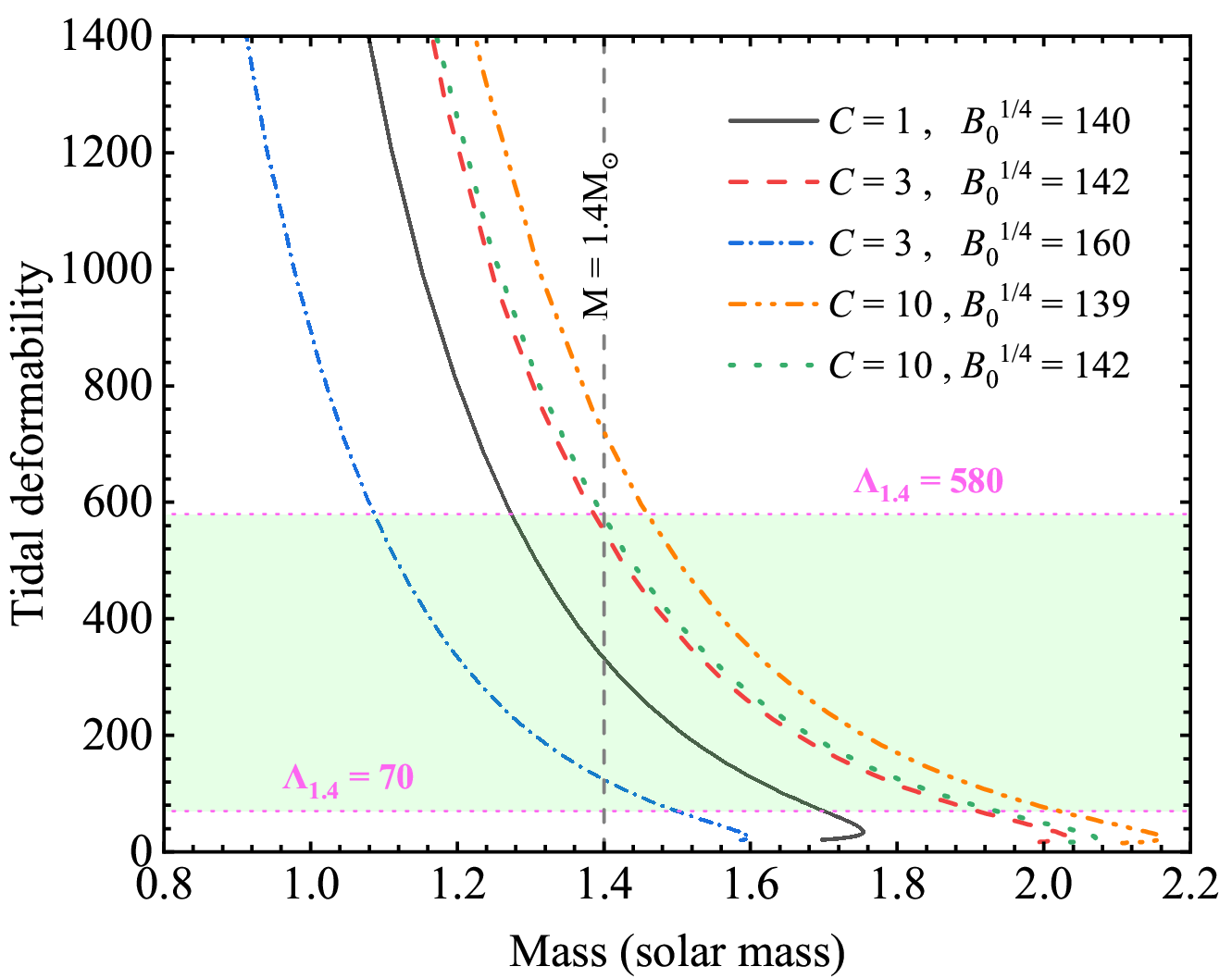}\\
  \caption{Tidal deformability of $ud$ quark stars as a function of the star mass for the same parameter sets as Fig.~\ref{fig:NbEnergy}. The black shaded area represents the improved estimate of the tidal deformability $\tilde{\Lambda}_{1.4}=190^{+390}_{-120}$ for GW170817 event. }\label{Fig:TidalD}
\end{figure}

In Fig.~\ref{Fig:TidalD}, we plot the dimensionless tidal deformability as a function of the gravitational mass for $ud$ quark stars in the revised perturbative QCD model with different parameter sets. The vertical dashed line marked a mass of $1.4~M_{\odot}$, and the shaded area represents the observational constraint on tidal deformability, $\tilde{\Lambda}_{1.4}=70-580$, derived from the LIGO-Virgo analysis of the GW170817 event. 
It is evident that the tidal deformability of $ud$ quark stars increases markedly as their mass decreases, indicating that lighter stars are more susceptible to deformation than heavier ones. The parameter set for the orange line, ($C,~B_0^{1/4}$), is taken from Region II, while all other curves are from Region III. Consequently, the intersection of the orange line with the vertical dashed line falls outside the shaded area, whereas the intersections for all other curves lie within it.
The distinction between parameter sets is crucial. The orange curve, corresponding to a set from Region II, behaves differently from all others (Region III). While the latter intersect the vertical $1.4 \, M_{\odot}$ line within the allowed shaded region, the orange curve's intersection lies outside it. This behavior is linked to the structural properties of the stars. As shown in the mass-radius relations (Fig. \ref{fig:MassRadius}), parameter sets that form more massive stars also do so with a larger radius. This structural difference directly impacts their tidal response. For a $1.4 \, M_{\odot}$ star, models with a larger maximum mass radius consequently exhibit a larger $\tilde{\Lambda}_{1.4}$, as their curves shift upward to intersect the $1.4 \, M_{\odot}$ line at a higher value.

\begin{table}
\caption{\label{table:Mmax} 
Maximum masses of $ud$ quark stars for several selected typical parameter sets. Note that both the maximum mass $ M_\text{max} $ and the radius are expressed in units of $ M_{\odot} $ and km, respectively. Meanwhile, the fourth root of the constant vacuum pressure $ B_0^{1/4} $ is given in units of MeV.
}
\setlength{\tabcolsep}{0.9pt}
\renewcommand\arraystretch{1.7}
\begin{ruledtabular}
\vspace{+0.1cm}
\begin{tabular*}{\hsize}{@{}@{\extracolsep{\fill}}c|ccc|cc@{}}
Parameters     &      &  Region~ III  &        &~~~~~~Region&II~~~~~~~\\
 \hline
($C,~B_0^{1/4}$) ~  &  (3, 142)    &  (10, 142)  &  (20, 142)  &  (10, 139)  &  (20, 139)   \\ 
$M_\text{max}$  &  2.03    &  2.07  &  2.08   &  2.16   &  2.17  \\
$R$  &  11.22   &  11.35  &  11.36   &   11.83  &  11.86  \\
\end{tabular*}
\end{ruledtabular}
\vspace{-0.5cm}
\end{table}

Very recently, the determination of the gravitational mass of the compact
star PSR J0740+6620 has been updated to $2.08^{+0.07}_{-0.07}$~$M_{\odot}$~\cite{Fonseca:2021wxt}. %
Moreover, two more massive compact stars, MSP J0740+6620~\cite{NANOGrav:2019jur} and PSR J2215+5135~\cite{Linares:2018ppq}, have been measured to be $2.14 \pm_{0.09}^{0.10}~M_{\odot}$ and $2.27_{-0.15}^{+0.17}~ M_{\odot}$, respectively. 
In Table~\ref{table:Mmax}, we summarize the results obtained from several selected  typical boundary parameters within the parameter range allowed by the revised perturbative QCD model shown in Fig.~\ref{fig:C0B0}, i.e., regions II and III, and calculated the corresponding maximum masses of the compact stars.
For the three sets of parameters in region III, we selected them at the lower boundary of region III in order to achieve the maximum stellar mass. These parameter sets ensure that the calculated tidal deformability of a 1.4 $M_\odot$ compact star is consistent with the observational results from the GW170817 event. Additionally, these parameter sets share a notable feature: the value of $B_0^{1/4}$ is fixed at 142 MeV, while the value of $C$ increases from 3 to 20. According to the calculation results, under the condition that $B_0$ remains constant, regardless of the value of $C$, the obtained maximum masses of the compact stars corresponding to these three parameter sets are all approximately 2 $M_\odot$, with differences among the three maximum masses not exceeding 3\%. Furthermore, as $C$ continues to increase, the calculated stellar masses gradually approach the saturation value of the maximum mass, which is confirmed to be 2.08 $M_\odot$.

The last two groups of parameters on the right side of Table~\ref{table:Mmax} are $(C,~B_0^{1/4}/\text{MeV}) = $(10, 139) and (20, 139), taken from the bottom of the allowed parameter region II. 
Although the $C$ values of these two parameter sets differ by 10, the corresponding maximum masses of $ud$ quark stars are both approximately 2.17 $M_\odot$. It is easy to verify that as $C$ increases further, the calculated maximum mass of quark stars will not exceed 2.17 $M_\odot$. 
From the above discussions, the following conclusions can be drawn: For the parameters in region II, which require only that the energy per baryon exceeds 930 MeV without imposing the tidal deformation constraints from the GW170817 event, the maximum mass of a quark star allowed by the revised perturbative QCD model is 2.17 $M_\odot$. However, if both conditions are simultaneously satisfied, the maximum mass allowed by the revised perturbative QCD model decreases to 2.07$M_\odot$. This result suggests that some compact objects observed in recent astronomical studies, such as an object with a mass of approximately 2.6$M_\odot$, are likely not quark stars composed of $ud$ quark matter, according to the revised perturbative QCD model.

\section{Conclusions}  \label{sec:CONCLUSION}

The conventional perturbative QCD model provides a precise description of QCD matter at high densities. However, it encounters thermodynamic consistency issues at low densities, 
which become 
more pronounced as the density decreases. To address this issue, a correction term, determined by the Cauchy criterion, is introduced into the thermodynamic potential density of the system. This modification effectively resolves the thermodynamic inconsistencies of the conventional perturbative QCD model. In the revised model, the energy minimum of the system locates exactly at the zero pressure point, fulfilling the thermodynamic consistency requirements of phenomenological models.

Within the framework of the revised perturbative QCD model, we have constrained the stability window of $ud$ quark matter by combining astrophysical observational data, in particular the constraints from standard nuclear physics, and the measured gravitational mass of PSR J0740$+$6620~\cite{Fonseca:2021wxt} 
with a lower limit of $2.01~M_{\odot}$ and the tidal deformability $\tilde{\Lambda}_{1.4}=190^{+390}_{-120}$ from the GW170817 event. Our study reveals that, on the $C-B_0^{1/4}$ parameter plane, as the parameter $C$ increases from small values to 1, the corresponding $B_0$ values of multiple constraint curves rise rapidly before stabilizing. With further increases in $C$, these curves saturate, signaling entry into a stable parameter regime. 
In other words, when $B_0$ remains constant and the parameter $C$ is greater than 1, the tidal deformability and maximum mass of the compact star remain almost constant as 
$C$ increases. This suggests that, these properties are less sensitive to $C$ and more dependent on $B_0$ in this parameter region. Additionally, we find that smaller values of $B_0$ and/or larger values of $C$ lead to larger maximum stellar masses and tidal deformability.

Through our analysis of the allowed parameter space, we find that when the energy per baryon of $ud$ quark matter does not conflict with the constraints from standard nuclear physics, i.e. $\epsilon_{ud}\geq 930~\text{MeV}$, the maximum mass of an $ud$ quark star allowed by the revised perturbative QCD model is 2.17$M_{\odot}$. However, when further considering the tidal deformability values measured in the GW170817 gravitational wave event $\tilde{\Lambda}_{1.4}=190^{+390}_{-120}$, the maximum mass of an $ud$ quark star allowed by the revised perturbative QCD model decreases to 2.08$M_{\odot}$.  
Consequently, the revised perturbative QCD model indicates that the compact object observed in GW190814, with a mass of $2.59^{+0.08}_{-0.09}~M_{\odot}$, is unlikely to be an $ud$ quark star.

Last but not least, we would like to emphasize that the conclusions presented above are mainly drawn under the assumption that $\epsilon_{ud}\geq 930~\text{MeV}$. 
However, an alternative viewpoint, as presented in Refs.~\cite{Cao:2020zxi,Holdom:2017gdc}, posits that $ud$ quark matter could be more stable than iron nuclei. Under this assumption, the black shaded region in the lower-right corner of Fig.~\ref{fig:C0B0} would become a viable parameter space. Within this region, it would be possible to form compact objects with masses significantly exceeding twice the solar mass, such as the secondary component of GW190814, with a reported mass of 2.50$-$2.67$M_{\odot}$.  
Nevertheless, this scenario is in tension with the observational data from the GW170817 event.

An extension of this work is that we can apply the revised perturbative QCD model to investigate the properties of QCD matter under finite isospin chemical potentials~\cite{Son:2000xc}. In this scenario, a notable advantage is that lattice simulations do not suffer from sign problems under such conditions, making them a reliable tool for comparison~\cite{Kogut:2002zg}. As a result, the predictions from perturbative QCD can be directly compared with lattice data, or conversely, lattice results can be used to constrain the parameter space of the revised perturbative QCD model. We are aware that in the ratio of the energy density of isospin-asymmetric strongly interacting matter to its Stefan-Boltzmann limit, a distinct peak structure emerges as the isospin chemical potential varies, as observed in both lattice simulations and chiral perturbation theory. However, the traditional perturbative QCD model fails to reproduce this peak. We anticipate that the revised perturbative QCD model will successfully capture this feature, leading to a deeper understanding of QCD matter under isospin asymmetry.
Moreover, this line of research will enable a more precise determination of the model parameters.
We plan to present the results of this study in the near future. Finally, motivated by recent developments in Bayesian inference applied to dense-matter physics, we also plan to conduct a Bayesian parameter-space analysis to systematically quantify the uncertainties and correlations in $C$ and $B_0$, thus enhancing the statistical robustness and interpretative strength of future astrophysical applications.


\section*{Acknowledgments}

ZYL thanks Dr.~Xun Chen and Prof.~Guang-Xiong Peng for useful discussions. 
This work is supported in part by the National Natural Science Foundation of China 
(Grants No.~12205093, No.~12404240, No.~12375045, and No.~12405054), the Hunan Provincial Natural Science Foundation of China (Grants No.~2021JJ40188 and No.~2024JJ6210), and the Scientific Research Fund of Hunan Provincial Education Department of China (Grants No.~19C0772 and No.~21A0297).

\section*{Data availability}

No data were created or analyzed in this study.


\bibliography{Ref}

\begin{thebibliography}{97}%
\makeatletter
\providecommand \@ifxundefined [1]{%
 \@ifx{#1\undefined}
}%
\providecommand \@ifnum [1]{%
 \ifnum #1\expandafter \@firstoftwo
 \else \expandafter \@secondoftwo
 \fi
}%
\providecommand \@ifx [1]{%
 \ifx #1\expandafter \@firstoftwo
 \else \expandafter \@secondoftwo
 \fi
}%
\providecommand \natexlab [1]{#1}%
\providecommand \enquote  [1]{``#1''}%
\providecommand \bibnamefont  [1]{#1}%
\providecommand \bibfnamefont [1]{#1}%
\providecommand \citenamefont [1]{#1}%
\providecommand \href@noop [0]{\@secondoftwo}%
\providecommand \href [0]{\begingroup \@sanitize@url \@href}%
\providecommand \@href[1]{\@@startlink{#1}\@@href}%
\providecommand \@@href[1]{\endgroup#1\@@endlink}%
\providecommand \@sanitize@url [0]{\catcode `\\12\catcode `\$12\catcode
  `\&12\catcode `\#12\catcode `\^12\catcode `\_12\catcode `\%12\relax}%
\providecommand \@@startlink[1]{}%
\providecommand \@@endlink[0]{}%
\providecommand \url  [0]{\begingroup\@sanitize@url \@url }%
\providecommand \@url [1]{\endgroup\@href {#1}{\urlprefix }}%
\providecommand \urlprefix  [0]{URL }%
\providecommand \Eprint [0]{\href }%
\providecommand \doibase [0]{http://dx.doi.org/}%
\providecommand \selectlanguage [0]{\@gobble}%
\providecommand \bibinfo  [0]{\@secondoftwo}%
\providecommand \bibfield  [0]{\@secondoftwo}%
\providecommand \translation [1]{[#1]}%
\providecommand \BibitemOpen [0]{}%
\providecommand \bibitemStop [0]{}%
\providecommand \bibitemNoStop [0]{.\EOS\space}%
\providecommand \EOS [0]{\spacefactor3000\relax}%
\providecommand \BibitemShut  [1]{\csname bibitem#1\endcsname}%
\let\auto@bib@innerbib\@empty
\bibitem [{\citenamefont {Abbott}\ \emph {et~al.}(2017)\citenamefont {Abbott}
  \emph {et~al.}}]{LIGOScientific:2017vwq}%
  \BibitemOpen
  \bibfield  {author} {\bibinfo {author} {\bibfnamefont {B.~P.}\ \bibnamefont
  {Abbott}} \emph {et~al.} (\bibinfo {collaboration} {LIGO Scientific and Virgo
  Collaborations}),\ }\bibfield  {title} {\enquote {\bibinfo {title}
  {{{GW170817}}: {{Observation}} of gravitational waves from a binary neutron
  star inspiral},}\ }\href {\doibase 10.1103/PhysRevLett.119.161101} {\bibfield
   {journal} {\bibinfo  {journal} {Phys. Rev. Lett.}\ }\textbf {\bibinfo
  {volume} {119}},\ \bibinfo {pages} {161101} (\bibinfo {year} {2017})},\
  \Eprint {http://arxiv.org/abs/1710.05832} {arXiv:1710.05832 [gr-qc]}
  \BibitemShut {NoStop}%
\bibitem [{\citenamefont {Abbott}\ \emph {et~al.}(2020)\citenamefont {Abbott}
  \emph {et~al.}}]{LIGOScientific:2020zkf}%
  \BibitemOpen
  \bibfield  {author} {\bibinfo {author} {\bibfnamefont {R.}~\bibnamefont
  {Abbott}} \emph {et~al.} (\bibinfo {collaboration} {LIGO Scientific,
  Virgo}),\ }\bibfield  {title} {\enquote {\bibinfo {title} {{{GW190814}}:
  {{Gravitational}} waves from the coalescence of a 23 solar mass black hole
  with a 2.6 solar mass compact object},}\ }\href {\doibase
  10.3847/2041-8213/ab960f} {\bibfield  {journal} {\bibinfo  {journal}
  {Astrophys. J. Lett.}\ }\textbf {\bibinfo {volume} {896}},\ \bibinfo {pages}
  {L44} (\bibinfo {year} {2020})},\ \Eprint {http://arxiv.org/abs/2006.12611}
  {arXiv:2006.12611 [astro-ph.HE]} \BibitemShut {NoStop}%
\bibitem [{\citenamefont {Riley}\ \emph {et~al.}(2021)\citenamefont {Riley}
  \emph {et~al.}}]{Riley:2021pdl}%
  \BibitemOpen
  \bibfield  {author} {\bibinfo {author} {\bibfnamefont {Thomas~E.}\
  \bibnamefont {Riley}} \emph {et~al.},\ }\bibfield  {title} {\enquote
  {\bibinfo {title} {A {{NICER View}} of the {{Massive Pulsar PSR J0740}}+6620
  {{Informed}} by {{Radio Timing}} and {{XMM-Newton Spectroscopy}}},}\ }\href
  {\doibase 10.3847/2041-8213/ac0a81} {\bibfield  {journal} {\bibinfo
  {journal} {Astrophys. J. Lett.}\ }\textbf {\bibinfo {volume} {918}},\
  \bibinfo {pages} {L27} (\bibinfo {year} {2021})}\BibitemShut {NoStop}%
\bibitem [{\citenamefont {Romani}\ \emph {et~al.}(2022)\citenamefont {Romani},
  \citenamefont {Kandel}, \citenamefont {Filippenko}, \citenamefont {Brink},\
  and\ \citenamefont {Zheng}}]{Romani:2022jhd}%
  \BibitemOpen
  \bibfield  {author} {\bibinfo {author} {\bibfnamefont {Roger~W.}\
  \bibnamefont {Romani}}, \bibinfo {author} {\bibfnamefont {D.}~\bibnamefont
  {Kandel}}, \bibinfo {author} {\bibfnamefont {Alexei~V.}\ \bibnamefont
  {Filippenko}}, \bibinfo {author} {\bibfnamefont {Thomas~G.}\ \bibnamefont
  {Brink}}, \ and\ \bibinfo {author} {\bibfnamefont {WeiKang}\ \bibnamefont
  {Zheng}},\ }\bibfield  {title} {\enquote {\bibinfo {title} {{{PSR
  J0952}}{\textbackslash}ensuremath-0607: {{The Fastest}} and {{Heaviest Known
  Galactic Neutron Star}}},}\ }\href {\doibase 10.3847/2041-8213/ac8007}
  {\bibfield  {journal} {\bibinfo  {journal} {Astrophys. J. Lett.}\ }\textbf
  {\bibinfo {volume} {934}},\ \bibinfo {pages} {L17} (\bibinfo {year}
  {2022})}\BibitemShut {NoStop}%
\bibitem [{\citenamefont {Ferreira}\ and\ \citenamefont
  {Provid{\^e}ncia}(2021)}]{Ferreira:2021pni}%
  \BibitemOpen
  \bibfield  {author} {\bibinfo {author} {\bibfnamefont {M{\'a}rcio}\
  \bibnamefont {Ferreira}}\ and\ \bibinfo {author} {\bibfnamefont {Constan{\c
  c}a}\ \bibnamefont {Provid{\^e}ncia}},\ }\bibfield  {title} {\enquote
  {\bibinfo {title} {Constraints on high density equation of state from maximum
  neutron star mass},}\ }\href {\doibase 10.1103/PhysRevD.104.063006}
  {\bibfield  {journal} {\bibinfo  {journal} {Phys. Rev. D}\ }\textbf {\bibinfo
  {volume} {104}},\ \bibinfo {pages} {063006} (\bibinfo {year} {2021})},\
  \Eprint {http://arxiv.org/abs/2110.00305} {arXiv:2110.00305 [nucl-th]}
  \BibitemShut {NoStop}%
\bibitem [{\citenamefont {{Kanakis-Pegios}}\ \emph {et~al.}(2021)\citenamefont
  {{Kanakis-Pegios}}, \citenamefont {Koliogiannis},\ and\ \citenamefont
  {Moustakidis}}]{Kanakis-Pegios:2020kzp}%
  \BibitemOpen
  \bibfield  {author} {\bibinfo {author} {\bibfnamefont {A.}~\bibnamefont
  {{Kanakis-Pegios}}}, \bibinfo {author} {\bibfnamefont {P.~S.}\ \bibnamefont
  {Koliogiannis}}, \ and\ \bibinfo {author} {\bibfnamefont {{\relax Ch}.~C.}\
  \bibnamefont {Moustakidis}},\ }\bibfield  {title} {\enquote {\bibinfo {title}
  {Probing the {{Nuclear Equation}} of {{State}} from the {{Existence}} of a
  \${\textbackslash}sim 2.6 {{M}}\_{\o}dot\$ {{Neutron Star}}: {{The GW190814
  Puzzle}}},}\ }\href {\doibase 10.3390/sym13020183} {\bibfield  {journal}
  {\bibinfo  {journal} {Symmetry}\ }\textbf {\bibinfo {volume} {13}},\ \bibinfo
  {pages} {183} (\bibinfo {year} {2021})}\BibitemShut {NoStop}%
\bibitem [{\citenamefont {Yang}\ \emph {et~al.}(2020)\citenamefont {Yang},
  \citenamefont {PI}, \citenamefont {Zheng},\ and\ \citenamefont
  {Weber}}]{Yang:2019rxn}%
  \BibitemOpen
  \bibfield  {author} {\bibinfo {author} {\bibfnamefont {Shu-Hua}\ \bibnamefont
  {Yang}}, \bibinfo {author} {\bibfnamefont {Chun-Mei}\ \bibnamefont {PI}},
  \bibinfo {author} {\bibfnamefont {Xiao-Ping}\ \bibnamefont {Zheng}}, \ and\
  \bibinfo {author} {\bibfnamefont {Fridolin}\ \bibnamefont {Weber}},\
  }\bibfield  {title} {\enquote {\bibinfo {title} {Non-newtonian gravity in
  strange quark stars and constraints from the observations of {{PSR
  J0740}}+6620 and {{GW170817}}},}\ }\href {\doibase 10.3847/1538-4357/abb365}
  {\bibfield  {journal} {\bibinfo  {journal} {Astrophys. J.}\ }\textbf
  {\bibinfo {volume} {902}},\ \bibinfo {pages} {32} (\bibinfo {year} {2020})},\
  \Eprint {http://arxiv.org/abs/1909.00933} {arXiv:1909.00933 [astro-ph.HE]}
  \BibitemShut {NoStop}%
\bibitem [{\citenamefont {Zhu}\ \emph {et~al.}(2023)\citenamefont {Zhu},
  \citenamefont {Li},\ and\ \citenamefont {Liu}}]{Zhu:2022ibs}%
  \BibitemOpen
  \bibfield  {author} {\bibinfo {author} {\bibfnamefont {Zhenyu}\ \bibnamefont
  {Zhu}}, \bibinfo {author} {\bibfnamefont {Ang}\ \bibnamefont {Li}}, \ and\
  \bibinfo {author} {\bibfnamefont {Tong}\ \bibnamefont {Liu}},\ }\bibfield
  {title} {\enquote {\bibinfo {title} {A {{Bayesian Inference}} of a
  {{Relativistic Mean-field Model}} of {{Neutron Star Matter}} from
  {{Observations}} of {{NICER}} and {{GW170817}}/{{AT2017gfo}}},}\ }\href
  {\doibase 10.3847/1538-4357/acac1f} {\bibfield  {journal} {\bibinfo
  {journal} {Astrophys. J.}\ }\textbf {\bibinfo {volume} {943}},\ \bibinfo
  {pages} {163} (\bibinfo {year} {2023})}\BibitemShut {NoStop}%
\bibitem [{\citenamefont {Dexheimer}\ \emph {et~al.}(2021)\citenamefont
  {Dexheimer}, \citenamefont {Gomes}, \citenamefont {Kl{\"a}hn}, \citenamefont
  {Han},\ and\ \citenamefont {Salinas}}]{Dexheimer:2020rlp}%
  \BibitemOpen
  \bibfield  {author} {\bibinfo {author} {\bibfnamefont {V.}~\bibnamefont
  {Dexheimer}}, \bibinfo {author} {\bibfnamefont {R.~O.}\ \bibnamefont
  {Gomes}}, \bibinfo {author} {\bibfnamefont {T.}~\bibnamefont {Kl{\"a}hn}},
  \bibinfo {author} {\bibfnamefont {S.}~\bibnamefont {Han}}, \ and\ \bibinfo
  {author} {\bibfnamefont {M.}~\bibnamefont {Salinas}},\ }\bibfield  {title}
  {\enquote {\bibinfo {title} {{{GW190814}} as a massive rapidly rotating
  neutron star with exotic degrees of freedom},}\ }\href {\doibase
  10.1103/PhysRevC.103.025808} {\bibfield  {journal} {\bibinfo  {journal}
  {Phys. Rev. C}\ }\textbf {\bibinfo {volume} {103}},\ \bibinfo {pages}
  {025808} (\bibinfo {year} {2021})}\BibitemShut {NoStop}%
\bibitem [{\citenamefont {Yang}\ \emph
  {et~al.}(2021{\natexlab{a}})\citenamefont {Yang}, \citenamefont {Pi},\ and\
  \citenamefont {Zheng}}]{Yang:2021bpe}%
  \BibitemOpen
  \bibfield  {author} {\bibinfo {author} {\bibfnamefont {Shu-Hua}\ \bibnamefont
  {Yang}}, \bibinfo {author} {\bibfnamefont {Chun-Mei}\ \bibnamefont {Pi}}, \
  and\ \bibinfo {author} {\bibfnamefont {Xiao-Ping}\ \bibnamefont {Zheng}},\
  }\bibfield  {title} {\enquote {\bibinfo {title} {Strange stars with a
  mirror-dark-matter core confronting with the observations of compact
  stars},}\ }\href {\doibase 10.1103/PhysRevD.104.083016} {\bibfield  {journal}
  {\bibinfo  {journal} {Phys. Rev. D}\ }\textbf {\bibinfo {volume} {104}},\
  \bibinfo {pages} {083016} (\bibinfo {year} {2021}{\natexlab{a}})},\ \Eprint
  {http://arxiv.org/abs/2103.05159} {arXiv:2103.05159 [astro-ph.HE]}
  \BibitemShut {NoStop}%
\bibitem [{\citenamefont {Nunes}\ \emph {et~al.}(2020)\citenamefont {Nunes},
  \citenamefont {Coelho},\ and\ \citenamefont {{de Araujo}}}]{Nunes:2020cuz}%
  \BibitemOpen
  \bibfield  {author} {\bibinfo {author} {\bibfnamefont {Rafael~C.}\
  \bibnamefont {Nunes}}, \bibinfo {author} {\bibfnamefont {Jaziel~G.}\
  \bibnamefont {Coelho}}, \ and\ \bibinfo {author} {\bibfnamefont {Jos{\'e}
  C.~N.}\ \bibnamefont {{de Araujo}}},\ }\bibfield  {title} {\enquote {\bibinfo
  {title} {Weighing massive neutron star with screening gravity: A look on
  {{PSR J0740}} + 6620 and {{GW190814}} secondary component},}\ }\href
  {\doibase 10.1140/epjc/s10052-020-08695-0} {\bibfield  {journal} {\bibinfo
  {journal} {Eur. Phys. J. C}\ }\textbf {\bibinfo {volume} {80}},\ \bibinfo
  {pages} {1115} (\bibinfo {year} {2020})}\BibitemShut {NoStop}%
\bibitem [{\citenamefont {Annala}\ \emph {et~al.}(2022)\citenamefont {Annala},
  \citenamefont {Gorda}, \citenamefont {Katerini}, \citenamefont {Kurkela},
  \citenamefont {N{\"a}ttil{\"a}}, \citenamefont {Paschalidis},\ and\
  \citenamefont {Vuorinen}}]{Annala:2021gom}%
  \BibitemOpen
  \bibfield  {author} {\bibinfo {author} {\bibfnamefont {Eemeli}\ \bibnamefont
  {Annala}}, \bibinfo {author} {\bibfnamefont {Tyler}\ \bibnamefont {Gorda}},
  \bibinfo {author} {\bibfnamefont {Evangelia}\ \bibnamefont {Katerini}},
  \bibinfo {author} {\bibfnamefont {Aleksi}\ \bibnamefont {Kurkela}}, \bibinfo
  {author} {\bibfnamefont {Joonas}\ \bibnamefont {N{\"a}ttil{\"a}}}, \bibinfo
  {author} {\bibfnamefont {Vasileios}\ \bibnamefont {Paschalidis}}, \ and\
  \bibinfo {author} {\bibfnamefont {Aleksi}\ \bibnamefont {Vuorinen}},\
  }\bibfield  {title} {\enquote {\bibinfo {title} {Multimessenger
  {{Constraints}} for {{Ultradense Matter}}},}\ }\href {\doibase
  10.1103/PhysRevX.12.011058} {\bibfield  {journal} {\bibinfo  {journal} {Phys.
  Rev. X}\ }\textbf {\bibinfo {volume} {12}},\ \bibinfo {pages} {011058}
  (\bibinfo {year} {2022})}\BibitemShut {NoStop}%
\bibitem [{\citenamefont {Lim}\ \emph {et~al.}(2021)\citenamefont {Lim},
  \citenamefont {Bhattacharya}, \citenamefont {Holt},\ and\ \citenamefont
  {Pati}}]{Lim:2020zvx}%
  \BibitemOpen
  \bibfield  {author} {\bibinfo {author} {\bibfnamefont {Yeunhwan}\
  \bibnamefont {Lim}}, \bibinfo {author} {\bibfnamefont {Anirban}\ \bibnamefont
  {Bhattacharya}}, \bibinfo {author} {\bibfnamefont {Jeremy~W.}\ \bibnamefont
  {Holt}}, \ and\ \bibinfo {author} {\bibfnamefont {Debdeep}\ \bibnamefont
  {Pati}},\ }\bibfield  {title} {\enquote {\bibinfo {title} {Radius and
  equation of state constraints from massive neutron stars and {{GW190814}}},}\
  }\href {\doibase 10.1103/PhysRevC.104.L032802} {\bibfield  {journal}
  {\bibinfo  {journal} {Phys. Rev. C}\ }\textbf {\bibinfo {volume} {104}},\
  \bibinfo {pages} {L032802} (\bibinfo {year} {2021})}\BibitemShut {NoStop}%
\bibitem [{\citenamefont {Oertel}\ \emph {et~al.}(2017)\citenamefont {Oertel},
  \citenamefont {Hempel}, \citenamefont {Kl{\"a}hn},\ and\ \citenamefont
  {Typel}}]{Oertel:2016bki}%
  \BibitemOpen
  \bibfield  {author} {\bibinfo {author} {\bibfnamefont {M.}~\bibnamefont
  {Oertel}}, \bibinfo {author} {\bibfnamefont {M.}~\bibnamefont {Hempel}},
  \bibinfo {author} {\bibfnamefont {T.}~\bibnamefont {Kl{\"a}hn}}, \ and\
  \bibinfo {author} {\bibfnamefont {S.}~\bibnamefont {Typel}},\ }\bibfield
  {title} {\enquote {\bibinfo {title} {Equations of state for supernovae and
  compact stars},}\ }\href {\doibase 10.1103/RevModPhys.89.015007} {\bibfield
  {journal} {\bibinfo  {journal} {Rev. Mod. Phys.}\ }\textbf {\bibinfo {volume}
  {89}},\ \bibinfo {pages} {015007} (\bibinfo {year} {2017})},\ \Eprint
  {http://arxiv.org/abs/1610.03361} {arXiv:1610.03361 [astro-ph.HE]}
  \BibitemShut {NoStop}%
\bibitem [{\citenamefont {Kumar}\ \emph {et~al.}(2024)\citenamefont {Kumar}
  \emph {et~al.}}]{MUSES:2023hyz}%
  \BibitemOpen
  \bibfield  {author} {\bibinfo {author} {\bibfnamefont {Rajesh}\ \bibnamefont
  {Kumar}} \emph {et~al.},\ }\bibfield  {title} {\enquote {\bibinfo {title}
  {Theoretical and experimental constraints for the equation of state of dense
  and hot matter},}\ }\href {\doibase 10.1007/s41114-024-00049-6} {\bibfield
  {journal} {\bibinfo  {journal} {Living Rev. Rel.}\ }\textbf {\bibinfo
  {volume} {27}},\ \bibinfo {pages} {3} (\bibinfo {year} {2024})}\BibitemShut
  {NoStop}%
\bibitem [{\citenamefont {Reinke~Pelicer}\ \emph {et~al.}(2025)\citenamefont
  {Reinke~Pelicer} \emph {et~al.}}]{ReinkePelicer:2025vuh}%
  \BibitemOpen
  \bibfield  {author} {\bibinfo {author} {\bibfnamefont {Mateus}\ \bibnamefont
  {Reinke~Pelicer}} \emph {et~al.},\ }\bibfield  {title} {\enquote {\bibinfo
  {title} {Building neutron stars with the {{MUSES}} calculation engine},}\
  }\href {\doibase 10.1103/PhysRevD.111.103037} {\bibfield  {journal} {\bibinfo
   {journal} {Phys. Rev. D}\ }\textbf {\bibinfo {volume} {111}},\ \bibinfo
  {pages} {103037} (\bibinfo {year} {2025})}\BibitemShut {NoStop}%
\bibitem [{\citenamefont {Abbott}\ \emph {et~al.}(2018)\citenamefont {Abbott}
  \emph {et~al.}}]{LIGOScientific:2018cki}%
  \BibitemOpen
  \bibfield  {author} {\bibinfo {author} {\bibfnamefont {B.~P.}\ \bibnamefont
  {Abbott}} \emph {et~al.} (\bibinfo {collaboration} {LIGO Scientific and Virgo
  Collaborations}),\ }\bibfield  {title} {\enquote {\bibinfo {title}
  {{{GW170817}}: {{Measurements}} of neutron star radii and equation of
  state},}\ }\href {\doibase 10.1103/PhysRevLett.121.161101} {\bibfield
  {journal} {\bibinfo  {journal} {Phys. Rev. Lett.}\ }\textbf {\bibinfo
  {volume} {121}},\ \bibinfo {pages} {161101} (\bibinfo {year} {2018})},\
  \Eprint {http://arxiv.org/abs/1805.11581} {arXiv:1805.11581 [gr-qc]}
  \BibitemShut {NoStop}%
\bibitem [{\citenamefont {Lu}\ \emph {et~al.}(2025)\citenamefont {Lu},
  \citenamefont {Gao}, \citenamefont {Luo}, \citenamefont {Chang},\ and\
  \citenamefont {Liu}}]{Lu:2025qyf}%
  \BibitemOpen
  \bibfield  {author} {\bibinfo {author} {\bibfnamefont {Yi}~\bibnamefont
  {Lu}}, \bibinfo {author} {\bibfnamefont {Fei}\ \bibnamefont {Gao}}, \bibinfo
  {author} {\bibfnamefont {Xiaofeng}\ \bibnamefont {Luo}}, \bibinfo {author}
  {\bibfnamefont {Lei}\ \bibnamefont {Chang}}, \ and\ \bibinfo {author}
  {\bibfnamefont {Yuxin}\ \bibnamefont {Liu}},\ }\bibfield  {title} {\enquote
  {\bibinfo {title} {Revealing the signal of {{QCD}} phase transition in
  heavy-ion collisions},}\ }\href {\doibase 10.1007/s11433-024-2619-7}
  {\bibfield  {journal} {\bibinfo  {journal} {Sci. China Phys. Mech. Astron.}\
  }\textbf {\bibinfo {volume} {68}},\ \bibinfo {pages} {251012} (\bibinfo
  {year} {2025})}\BibitemShut {NoStop}%
\bibitem [{\citenamefont {Aarts}\ \emph {et~al.}(2023)\citenamefont {Aarts}
  \emph {et~al.}}]{Aarts:2023vsf}%
  \BibitemOpen
  \bibfield  {author} {\bibinfo {author} {\bibfnamefont {Gert}\ \bibnamefont
  {Aarts}} \emph {et~al.},\ }\bibfield  {title} {\enquote {\bibinfo {title}
  {Phase {{Transitions}} in {{Particle Physics}}: {{Results}} and
  {{Perspectives}} from {{Lattice Quantum Chromo-Dynamics}}},}\ }\href
  {\doibase 10.1016/j.ppnp.2023.104070} {\bibfield  {journal} {\bibinfo
  {journal} {Prog. Part. Nucl. Phys.}\ }\textbf {\bibinfo {volume} {133}},\
  \bibinfo {pages} {104070} (\bibinfo {year} {2023})}\BibitemShut {NoStop}%
\bibitem [{\citenamefont {Roupas}\ \emph {et~al.}(2021)\citenamefont {Roupas},
  \citenamefont {Panotopoulos},\ and\ \citenamefont {Lopes}}]{Roupas:2020nua}%
  \BibitemOpen
  \bibfield  {author} {\bibinfo {author} {\bibfnamefont {Zacharias}\
  \bibnamefont {Roupas}}, \bibinfo {author} {\bibfnamefont {Grigoris}\
  \bibnamefont {Panotopoulos}}, \ and\ \bibinfo {author} {\bibfnamefont
  {Il{\'i}dio}\ \bibnamefont {Lopes}},\ }\bibfield  {title} {\enquote {\bibinfo
  {title} {{{QCD}} color superconductivity in compact stars: Color-flavor
  locked quark star candidate for the gravitational-wave signal
  {{GW190814}}},}\ }\href {\doibase 10.1103/PhysRevD.103.083015} {\bibfield
  {journal} {\bibinfo  {journal} {Phys. Rev. D}\ }\textbf {\bibinfo {volume}
  {103}},\ \bibinfo {pages} {083015} (\bibinfo {year} {2021})}\BibitemShut
  {NoStop}%
\bibitem [{\citenamefont {Yang}\ \emph {et~al.}(2023)\citenamefont {Yang},
  \citenamefont {Pi}, \citenamefont {Zheng},\ and\ \citenamefont
  {Weber}}]{Yang:2023haz}%
  \BibitemOpen
  \bibfield  {author} {\bibinfo {author} {\bibfnamefont {Shu-Hua}\ \bibnamefont
  {Yang}}, \bibinfo {author} {\bibfnamefont {Chun-Mei}\ \bibnamefont {Pi}},
  \bibinfo {author} {\bibfnamefont {Xiao-Ping}\ \bibnamefont {Zheng}}, \ and\
  \bibinfo {author} {\bibfnamefont {Fridolin}\ \bibnamefont {Weber}},\
  }\bibfield  {title} {\enquote {\bibinfo {title} {Confronting strange stars
  with compact-star observations and new physics},}\ }\href {\doibase
  10.3390/universe9050202} {\bibfield  {journal} {\bibinfo  {journal}
  {Universe}\ }\textbf {\bibinfo {volume} {9}},\ \bibinfo {pages} {202}
  (\bibinfo {year} {2023})},\ \Eprint {http://arxiv.org/abs/2304.09614}
  {arXiv:2304.09614 [astro-ph.HE]} \BibitemShut {NoStop}%
\bibitem [{\citenamefont {Yuan}\ \emph {et~al.}(2025)\citenamefont {Yuan},
  \citenamefont {Huang}, \citenamefont {Zhang}, \citenamefont {Zhou},\ and\
  \citenamefont {Xu}}]{Yuan:2024hge}%
  \BibitemOpen
  \bibfield  {author} {\bibinfo {author} {\bibfnamefont {Wen-Li}\ \bibnamefont
  {Yuan}}, \bibinfo {author} {\bibfnamefont {Chun}\ \bibnamefont {Huang}},
  \bibinfo {author} {\bibfnamefont {Chen}\ \bibnamefont {Zhang}}, \bibinfo
  {author} {\bibfnamefont {Enping}\ \bibnamefont {Zhou}}, \ and\ \bibinfo
  {author} {\bibfnamefont {Renxin}\ \bibnamefont {Xu}},\ }\bibfield  {title}
  {\enquote {\bibinfo {title} {Bayesian inference of strangeon matter using the
  measurements of {{PSR J0437-4715}} and {{GW190814}}},}\ }\href {\doibase
  10.1103/PhysRevD.111.063033} {\bibfield  {journal} {\bibinfo  {journal}
  {Phys. Rev. D}\ }\textbf {\bibinfo {volume} {111}},\ \bibinfo {pages}
  {063033} (\bibinfo {year} {2025})}\BibitemShut {NoStop}%
\bibitem [{\citenamefont {Tews}\ \emph {et~al.}(2018)\citenamefont {Tews},
  \citenamefont {Margueron},\ and\ \citenamefont {Reddy}}]{Tews:2018iwm}%
  \BibitemOpen
  \bibfield  {author} {\bibinfo {author} {\bibfnamefont {I.}~\bibnamefont
  {Tews}}, \bibinfo {author} {\bibfnamefont {J.}~\bibnamefont {Margueron}}, \
  and\ \bibinfo {author} {\bibfnamefont {S.}~\bibnamefont {Reddy}},\ }\bibfield
   {title} {\enquote {\bibinfo {title} {Critical examination of constraints on
  the equation of state of dense matter obtained from {{GW170817}}},}\ }\href
  {\doibase 10.1103/PhysRevC.98.045804} {\bibfield  {journal} {\bibinfo
  {journal} {Phys. Rev. C}\ }\textbf {\bibinfo {volume} {98}},\ \bibinfo
  {pages} {045804} (\bibinfo {year} {2018})}\BibitemShut {NoStop}%
\bibitem [{\citenamefont {Li}\ \emph {et~al.}(2021{\natexlab{a}})\citenamefont
  {Li}, \citenamefont {Miao}, \citenamefont {Han},\ and\ \citenamefont
  {Zhang}}]{Li:2021crp}%
  \BibitemOpen
  \bibfield  {author} {\bibinfo {author} {\bibfnamefont {Ang}\ \bibnamefont
  {Li}}, \bibinfo {author} {\bibfnamefont {Zhiqiang}\ \bibnamefont {Miao}},
  \bibinfo {author} {\bibfnamefont {Sophia}\ \bibnamefont {Han}}, \ and\
  \bibinfo {author} {\bibfnamefont {Bing}\ \bibnamefont {Zhang}},\ }\bibfield
  {title} {\enquote {\bibinfo {title} {Constraints on the {{Maximum Mass}} of
  {{Neutron Stars}} with a {{Quark Core}} from {{GW170817}} and {{NICER PSR
  J0030}}+0451 {{Data}}},}\ }\href {\doibase 10.3847/1538-4357/abf355}
  {\bibfield  {journal} {\bibinfo  {journal} {ApJ}\ }\textbf {\bibinfo {volume}
  {913}},\ \bibinfo {pages} {27} (\bibinfo {year}
  {2021}{\natexlab{a}})}\BibitemShut {NoStop}%
\bibitem [{\citenamefont {Oikonomou}\ and\ \citenamefont
  {Moustakidis}(2023)}]{Oikonomou:2023otn}%
  \BibitemOpen
  \bibfield  {author} {\bibinfo {author} {\bibfnamefont {P.~T.}\ \bibnamefont
  {Oikonomou}}\ and\ \bibinfo {author} {\bibfnamefont {{\relax Ch}.~C.}\
  \bibnamefont {Moustakidis}},\ }\bibfield  {title} {\enquote {\bibinfo {title}
  {Color-flavor locked quark stars in light of the compact object in the {{HESS
  J1731-347}} and the {{GW190814}} event},}\ }\href {\doibase
  10.1103/PhysRevD.108.063010} {\bibfield  {journal} {\bibinfo  {journal}
  {Phys. Rev. D}\ }\textbf {\bibinfo {volume} {108}},\ \bibinfo {pages}
  {063010} (\bibinfo {year} {2023})}\BibitemShut {NoStop}%
\bibitem [{\citenamefont {Pi}\ and\ \citenamefont {Yang}(2022)}]{Pi:2022pjs}%
  \BibitemOpen
  \bibfield  {author} {\bibinfo {author} {\bibfnamefont {Chun-Mei}\
  \bibnamefont {Pi}}\ and\ \bibinfo {author} {\bibfnamefont {Shu-Hua}\
  \bibnamefont {Yang}},\ }\bibfield  {title} {\enquote {\bibinfo {title}
  {Non-{{Newtonian}} gravity in strange stars and constraints from the
  observations of compact stars},}\ }\href {\doibase
  10.1016/j.newast.2021.101670} {\bibfield  {journal} {\bibinfo  {journal} {New
  Astron.}\ }\textbf {\bibinfo {volume} {90}},\ \bibinfo {pages} {101670}
  (\bibinfo {year} {2022})}\BibitemShut {NoStop}%
\bibitem [{\citenamefont {Tangphati}\ \emph
  {et~al.}(2024{\natexlab{a}})\citenamefont {Tangphati}, \citenamefont
  {Banerjee}, \citenamefont {Sakall{\i}},\ and\ \citenamefont
  {Pradhan}}]{Tangphati:2024atj}%
  \BibitemOpen
  \bibfield  {author} {\bibinfo {author} {\bibfnamefont {Takol}\ \bibnamefont
  {Tangphati}}, \bibinfo {author} {\bibfnamefont {Ayan}\ \bibnamefont
  {Banerjee}}, \bibinfo {author} {\bibfnamefont {{\.I}zzet}\ \bibnamefont
  {Sakall{\i}}}, \ and\ \bibinfo {author} {\bibfnamefont {Anirudh}\
  \bibnamefont {Pradhan}},\ }\bibfield  {title} {\enquote {\bibinfo {title}
  {Quark stars in {{Rastall}} gravity with recent astrophysical
  observations},}\ }\href {\doibase 10.1016/j.cjph.2024.03.045} {\bibfield
  {journal} {\bibinfo  {journal} {Chin. J. Phys.}\ }\textbf {\bibinfo {volume}
  {90}},\ \bibinfo {pages} {422--433} (\bibinfo {year}
  {2024}{\natexlab{a}})}\BibitemShut {NoStop}%
\bibitem [{\citenamefont {Brodie}\ and\ \citenamefont
  {Haber}(2023)}]{Brodie:2023pjw}%
  \BibitemOpen
  \bibfield  {author} {\bibinfo {author} {\bibfnamefont {Liam}\ \bibnamefont
  {Brodie}}\ and\ \bibinfo {author} {\bibfnamefont {Alexander}\ \bibnamefont
  {Haber}},\ }\bibfield  {title} {\enquote {\bibinfo {title} {Nuclear and
  hybrid equations of state in light of the low-mass compact star in {{HESS
  J1731-347}}},}\ }\href {\doibase 10.1103/PhysRevC.108.025806} {\bibfield
  {journal} {\bibinfo  {journal} {Phys. Rev. C}\ }\textbf {\bibinfo {volume}
  {108}},\ \bibinfo {pages} {025806} (\bibinfo {year} {2023})}\BibitemShut
  {NoStop}%
\bibitem [{\citenamefont {Pang}\ \emph {et~al.}(2021)\citenamefont {Pang},
  \citenamefont {Tews}, \citenamefont {Coughlin}, \citenamefont {Bulla},
  \citenamefont {Van Den~Broeck},\ and\ \citenamefont
  {Dietrich}}]{Pang:2021jta}%
  \BibitemOpen
  \bibfield  {author} {\bibinfo {author} {\bibfnamefont {Peter T.~H.}\
  \bibnamefont {Pang}}, \bibinfo {author} {\bibfnamefont {Ingo}\ \bibnamefont
  {Tews}}, \bibinfo {author} {\bibfnamefont {Michael~W.}\ \bibnamefont
  {Coughlin}}, \bibinfo {author} {\bibfnamefont {Mattia}\ \bibnamefont
  {Bulla}}, \bibinfo {author} {\bibfnamefont {Chris}\ \bibnamefont {Van
  Den~Broeck}}, \ and\ \bibinfo {author} {\bibfnamefont {Tim}\ \bibnamefont
  {Dietrich}},\ }\bibfield  {title} {\enquote {\bibinfo {title} {Nuclear
  {{Physics Multimessenger Astrophysics Constraints}} on the {{Neutron Star
  Equation}} of {{State}}: {{Adding NICER}}'s {{PSR J0740}}+6620
  {{Measurement}}},}\ }\href {\doibase 10.3847/1538-4357/ac19ab} {\bibfield
  {journal} {\bibinfo  {journal} {Astrophys. J.}\ }\textbf {\bibinfo {volume}
  {922}},\ \bibinfo {pages} {14} (\bibinfo {year} {2021})}\BibitemShut
  {NoStop}%
\bibitem [{\citenamefont {Doroshenko}\ \emph {et~al.}(2022)\citenamefont
  {Doroshenko}, \citenamefont {Suleimanov}, \citenamefont {P{\"u}hlhofer},\
  and\ \citenamefont {Santangelo}}]{Doroshenko:2022nwp}%
  \BibitemOpen
  \bibfield  {author} {\bibinfo {author} {\bibfnamefont {Victor}\ \bibnamefont
  {Doroshenko}}, \bibinfo {author} {\bibfnamefont {Valery}\ \bibnamefont
  {Suleimanov}}, \bibinfo {author} {\bibfnamefont {Gerd}\ \bibnamefont
  {P{\"u}hlhofer}}, \ and\ \bibinfo {author} {\bibfnamefont {Andrea}\
  \bibnamefont {Santangelo}},\ }\bibfield  {title} {\enquote {\bibinfo {title}
  {A strangely light neutron star within a supernova remnant},}\ }\href
  {\doibase 10.1038/s41550-022-01800-1} {\bibfield  {journal} {\bibinfo
  {journal} {Nature Astron.}\ }\textbf {\bibinfo {volume} {6}},\ \bibinfo
  {pages} {1444--1451} (\bibinfo {year} {2022})}\BibitemShut {NoStop}%
\bibitem [{\citenamefont {Pal}\ \emph {et~al.}(2025)\citenamefont {Pal},
  \citenamefont {Podder},\ and\ \citenamefont {Chaudhuri}}]{Pal:2025skz}%
  \BibitemOpen
  \bibfield  {author} {\bibinfo {author} {\bibfnamefont {Suman}\ \bibnamefont
  {Pal}}, \bibinfo {author} {\bibfnamefont {Soumen}\ \bibnamefont {Podder}}, \
  and\ \bibinfo {author} {\bibfnamefont {Gargi}\ \bibnamefont {Chaudhuri}},\
  }\bibfield  {title} {\enquote {\bibinfo {title} {Is the central compact
  object in {{HESS J1731-347}} a hybrid star with a quark core? {{An}} analysis
  with the constant speed of sound parametrization},}\ }\href {\doibase
  10.3847/1538-4357/adbc6b} {\bibfield  {journal} {\bibinfo  {journal}
  {Astrophys. J.}\ }\textbf {\bibinfo {volume} {983}},\ \bibinfo {pages} {24}
  (\bibinfo {year} {2025})}\BibitemShut {NoStop}%
\bibitem [{\citenamefont {Ayriyan}\ \emph {et~al.}(2025)\citenamefont
  {Ayriyan}, \citenamefont {Blaschke}, \citenamefont {Carlomagno},
  \citenamefont {Contrera},\ and\ \citenamefont {Grunfeld}}]{Ayriyan:2024zfw}%
  \BibitemOpen
  \bibfield  {author} {\bibinfo {author} {\bibfnamefont {Alexander}\
  \bibnamefont {Ayriyan}}, \bibinfo {author} {\bibfnamefont {David}\
  \bibnamefont {Blaschke}}, \bibinfo {author} {\bibfnamefont {Juan~Pablo}\
  \bibnamefont {Carlomagno}}, \bibinfo {author} {\bibfnamefont {Gustavo~A.}\
  \bibnamefont {Contrera}}, \ and\ \bibinfo {author} {\bibfnamefont
  {Ana~Gabriela}\ \bibnamefont {Grunfeld}},\ }\bibfield  {title} {\enquote
  {\bibinfo {title} {Bayesian {{Analysis}} of {{Hybrid Neutron Star EOS
  Constraints Within}} an {{Instantaneous Nonlocal Chiral Quark Matter
  Model}}},}\ }\href {\doibase 10.3390/universe11050141} {\bibfield  {journal}
  {\bibinfo  {journal} {Universe}\ }\textbf {\bibinfo {volume} {11}},\ \bibinfo
  {pages} {141} (\bibinfo {year} {2025})}\BibitemShut {NoStop}%
\bibitem [{\citenamefont {Pal}\ and\ \citenamefont
  {Chaudhuri}(2024)}]{Pal:2024afl}%
  \BibitemOpen
  \bibfield  {author} {\bibinfo {author} {\bibfnamefont {Suman}\ \bibnamefont
  {Pal}}\ and\ \bibinfo {author} {\bibfnamefont {Gargi}\ \bibnamefont
  {Chaudhuri}},\ }\bibfield  {title} {\enquote {\bibinfo {title} {Effect of
  dark matter interaction on hybrid star in the light of the recent
  astrophysical observations},}\ }\href {\doibase
  10.1088/1475-7516/2024/10/064} {\bibfield  {journal} {\bibinfo  {journal}
  {JCAP}\ }\textbf {\bibinfo {volume} {10}},\ \bibinfo {pages} {064} (\bibinfo
  {year} {2024})}\BibitemShut {NoStop}%
\bibitem [{\citenamefont {Mariani}\ \emph {et~al.}(2022)\citenamefont
  {Mariani}, \citenamefont {Tonetto}, \citenamefont {Rodr{\'i}guez},
  \citenamefont {Celi}, \citenamefont {{Ranea-Sandoval}}, \citenamefont
  {Orsaria},\ and\ \citenamefont {P{\'e}rez~Mart{\'i}nez}}]{Mariani:2022xek}%
  \BibitemOpen
  \bibfield  {author} {\bibinfo {author} {\bibfnamefont {Mauro}\ \bibnamefont
  {Mariani}}, \bibinfo {author} {\bibfnamefont {Lucas}\ \bibnamefont
  {Tonetto}}, \bibinfo {author} {\bibfnamefont {M.~Camila}\ \bibnamefont
  {Rodr{\'i}guez}}, \bibinfo {author} {\bibfnamefont {Marcos~O.}\ \bibnamefont
  {Celi}}, \bibinfo {author} {\bibfnamefont {Ignacio~F.}\ \bibnamefont
  {{Ranea-Sandoval}}}, \bibinfo {author} {\bibfnamefont {Milva~G.}\
  \bibnamefont {Orsaria}}, \ and\ \bibinfo {author} {\bibfnamefont {Aurora}\
  \bibnamefont {P{\'e}rez~Mart{\'i}nez}},\ }\bibfield  {title} {\enquote
  {\bibinfo {title} {Oscillating magnetized hybrid stars under the magnifying
  glass of multimessenger observations},}\ }\href {\doibase
  10.1093/mnras/stac546} {\bibfield  {journal} {\bibinfo  {journal} {Mon. Not.
  Roy. Astron. Soc.}\ }\textbf {\bibinfo {volume} {512}},\ \bibinfo {pages}
  {517--534} (\bibinfo {year} {2022})}\BibitemShut {NoStop}%
\bibitem [{\citenamefont {Ferreira}\ \emph {et~al.}(2021)\citenamefont
  {Ferreira}, \citenamefont {C{\^a}mara~Pereira},\ and\ \citenamefont
  {Provid{\^e}ncia}}]{Ferreira:2021osk}%
  \BibitemOpen
  \bibfield  {author} {\bibinfo {author} {\bibfnamefont {M{\'a}rcio}\
  \bibnamefont {Ferreira}}, \bibinfo {author} {\bibfnamefont {Renan}\
  \bibnamefont {C{\^a}mara~Pereira}}, \ and\ \bibinfo {author} {\bibfnamefont
  {Constan{\c c}a}\ \bibnamefont {Provid{\^e}ncia}},\ }\bibfield  {title}
  {\enquote {\bibinfo {title} {Hybrid stars with large strange quark cores
  constrained by {{GW170817}}},}\ }\href {\doibase 10.1103/PhysRevD.103.123020}
  {\bibfield  {journal} {\bibinfo  {journal} {Phys. Rev. D}\ }\textbf {\bibinfo
  {volume} {103}},\ \bibinfo {pages} {123020} (\bibinfo {year} {2021})},\
  \Eprint {http://arxiv.org/abs/2105.06239} {arXiv:2105.06239 [nucl-th]}
  \BibitemShut {NoStop}%
\bibitem [{\citenamefont {{Ranea-Sandoval}}\ \emph {et~al.}(2019)\citenamefont
  {{Ranea-Sandoval}}, \citenamefont {Orsaria}, \citenamefont {Malfatti},
  \citenamefont {Curin},\ and\ \citenamefont {{al}}}]{Ranea-Sandoval:2019miz}%
  \BibitemOpen
  \bibfield  {author} {\bibinfo {author} {\bibfnamefont {Ignacio~Francisco}\
  \bibnamefont {{Ranea-Sandoval}}}, \bibinfo {author} {\bibfnamefont
  {Milva~Gabriela}\ \bibnamefont {Orsaria}}, \bibinfo {author} {\bibfnamefont
  {Germ{\'a}n}\ \bibnamefont {Malfatti}}, \bibinfo {author} {\bibfnamefont
  {Daniela}\ \bibnamefont {Curin}}, \ and\ \bibinfo {author} {\bibfnamefont
  {et}~\bibnamefont {{al}}},\ }\bibfield  {title} {\enquote {\bibinfo {title}
  {Effects of hadron-quark phase transitions in hybrid stars within the {{NJL}}
  model},}\ }\href {\doibase 10.3390/sym11030425} {\bibfield  {journal}
  {\bibinfo  {journal} {Symmetry}\ }\textbf {\bibinfo {volume} {11}},\ \bibinfo
  {pages} {425} (\bibinfo {year} {2019})},\ \Eprint
  {http://arxiv.org/abs/1903.11974} {arXiv:1903.11974 [nucl-th]} \BibitemShut
  {NoStop}%
\bibitem [{\citenamefont {Li}\ \emph {et~al.}(2021{\natexlab{b}})\citenamefont
  {Li}, \citenamefont {Sedrakian},\ and\ \citenamefont {Alford}}]{Li:2021sxb}%
  \BibitemOpen
  \bibfield  {author} {\bibinfo {author} {\bibfnamefont {Jia~Jie}\ \bibnamefont
  {Li}}, \bibinfo {author} {\bibfnamefont {Armen}\ \bibnamefont {Sedrakian}}, \
  and\ \bibinfo {author} {\bibfnamefont {Mark}\ \bibnamefont {Alford}},\
  }\bibfield  {title} {\enquote {\bibinfo {title} {Relativistic hybrid stars in
  light of the {{NICER PSR J0740}}+6620 radius measurement},}\ }\href {\doibase
  10.1103/PhysRevD.104.L121302} {\bibfield  {journal} {\bibinfo  {journal}
  {Phys. Rev. D}\ }\textbf {\bibinfo {volume} {104}},\ \bibinfo {pages}
  {L121302} (\bibinfo {year} {2021}{\natexlab{b}})}\BibitemShut {NoStop}%
\bibitem [{\citenamefont {Albino}\ \emph {et~al.}(2024)\citenamefont {Albino},
  \citenamefont {Malik}, \citenamefont {Ferreira},\ and\ \citenamefont
  {Provid{\^e}ncia}}]{Albino:2024ymc}%
  \BibitemOpen
  \bibfield  {author} {\bibinfo {author} {\bibfnamefont {Milena}\ \bibnamefont
  {Albino}}, \bibinfo {author} {\bibfnamefont {Tuhin}\ \bibnamefont {Malik}},
  \bibinfo {author} {\bibfnamefont {M{\'a}rcio}\ \bibnamefont {Ferreira}}, \
  and\ \bibinfo {author} {\bibfnamefont {Constan{\c c}a}\ \bibnamefont
  {Provid{\^e}ncia}},\ }\bibfield  {title} {\enquote {\bibinfo {title} {Hybrid
  star properties with the {{NJL}} and mean field approximation of {{QCD}}
  models: {{A Bayesian}} approach},}\ }\href {\doibase
  10.1103/PhysRevD.110.083037} {\bibfield  {journal} {\bibinfo  {journal}
  {Phys. Rev. D}\ }\textbf {\bibinfo {volume} {110}},\ \bibinfo {pages}
  {083037} (\bibinfo {year} {2024})}\BibitemShut {NoStop}%
\bibitem [{\citenamefont {Blaschke}\ \emph {et~al.}(2023)\citenamefont
  {Blaschke}, \citenamefont {Shukla}, \citenamefont {Ivanytskyi},\ and\
  \citenamefont {Liebing}}]{Blaschke:2022egm}%
  \BibitemOpen
  \bibfield  {author} {\bibinfo {author} {\bibfnamefont {David}\ \bibnamefont
  {Blaschke}}, \bibinfo {author} {\bibfnamefont {Udita}\ \bibnamefont
  {Shukla}}, \bibinfo {author} {\bibfnamefont {Oleksii}\ \bibnamefont
  {Ivanytskyi}}, \ and\ \bibinfo {author} {\bibfnamefont {Simon}\ \bibnamefont
  {Liebing}},\ }\bibfield  {title} {\enquote {\bibinfo {title} {Effect of color
  superconductivity on the mass of hybrid neutron stars in an effective model
  with perturbative {{QCD}} asymptotics},}\ }\href {\doibase
  10.1103/PhysRevD.107.063034} {\bibfield  {journal} {\bibinfo  {journal}
  {Phys. Rev. D}\ }\textbf {\bibinfo {volume} {107}},\ \bibinfo {pages}
  {063034} (\bibinfo {year} {2023})}\BibitemShut {NoStop}%
\bibitem [{\citenamefont {Tangphati}\ \emph
  {et~al.}(2024{\natexlab{b}})\citenamefont {Tangphati}, \citenamefont
  {Sakall{\i}}, \citenamefont {Banerjee},\ and\ \citenamefont
  {Ali}}]{Tangphati:2024ycu}%
  \BibitemOpen
  \bibfield  {author} {\bibinfo {author} {\bibfnamefont {Takol}\ \bibnamefont
  {Tangphati}}, \bibinfo {author} {\bibfnamefont {{\.I}zzet}\ \bibnamefont
  {Sakall{\i}}}, \bibinfo {author} {\bibfnamefont {Ayan}\ \bibnamefont
  {Banerjee}}, \ and\ \bibinfo {author} {\bibfnamefont {Akram}\ \bibnamefont
  {Ali}},\ }\bibfield  {title} {\enquote {\bibinfo {title} {Interacting quark
  star with pressure anisotropy and recent astrophysical observations},}\
  }\href {\doibase 10.1016/j.cjph.2024.07.019} {\bibfield  {journal} {\bibinfo
  {journal} {Chin. J. Phys.}\ }\textbf {\bibinfo {volume} {91}},\ \bibinfo
  {pages} {392--405} (\bibinfo {year} {2024}{\natexlab{b}})}\BibitemShut
  {NoStop}%
\bibitem [{\citenamefont {Xu}\ \emph {et~al.}(2022)\citenamefont {Xu},
  \citenamefont {Xia}, \citenamefont {Lu}, \citenamefont {Peng},\ and\
  \citenamefont {Zhao}}]{Xu:2022squ}%
  \BibitemOpen
  \bibfield  {author} {\bibinfo {author} {\bibfnamefont {Jian-Feng}\
  \bibnamefont {Xu}}, \bibinfo {author} {\bibfnamefont {Cheng-Jun}\
  \bibnamefont {Xia}}, \bibinfo {author} {\bibfnamefont {Zhen-Yan}\
  \bibnamefont {Lu}}, \bibinfo {author} {\bibfnamefont {Guang-Xiong}\
  \bibnamefont {Peng}}, \ and\ \bibinfo {author} {\bibfnamefont {Ya-Peng}\
  \bibnamefont {Zhao}},\ }\bibfield  {title} {\enquote {\bibinfo {title}
  {Symmetry energy of strange quark matter and tidal deformability of strange
  quark stars},}\ }\href {\doibase 10.1007/s41365-022-01130-x} {\bibfield
  {journal} {\bibinfo  {journal} {Nucl. Sci. Tech.}\ }\textbf {\bibinfo
  {volume} {33}},\ \bibinfo {pages} {143} (\bibinfo {year} {2022})}\BibitemShut
  {NoStop}%
\bibitem [{\citenamefont {Banerjee}\ \emph {et~al.}(2025)\citenamefont
  {Banerjee}, \citenamefont {Sakall{\i}}, \citenamefont {Dayanandan},\ and\
  \citenamefont {Pradhan}}]{Banerjee:2025ufb}%
  \BibitemOpen
  \bibfield  {author} {\bibinfo {author} {\bibfnamefont {Ayan}\ \bibnamefont
  {Banerjee}}, \bibinfo {author} {\bibfnamefont {{\.I}zzet}\ \bibnamefont
  {Sakall{\i}}}, \bibinfo {author} {\bibfnamefont {B.}~\bibnamefont
  {Dayanandan}}, \ and\ \bibinfo {author} {\bibfnamefont {Anirudh}\
  \bibnamefont {Pradhan}},\ }\bibfield  {title} {\enquote {\bibinfo {title}
  {Quark stars in f({{R}}, {{T}}) gravity: Mass-to-radius profiles and
  observational data},}\ }\href {\doibase 10.1088/1674-1137/ad86af} {\bibfield
  {journal} {\bibinfo  {journal} {Chin. Phys. C}\ }\textbf {\bibinfo {volume}
  {49}},\ \bibinfo {pages} {015102} (\bibinfo {year} {2025})}\BibitemShut
  {NoStop}%
\bibitem [{\citenamefont {Tangphati}\ \emph
  {et~al.}(2022{\natexlab{a}})\citenamefont {Tangphati}, \citenamefont {Karar},
  \citenamefont {Pradhan},\ and\ \citenamefont {Banerjee}}]{Tangphati:2022arm}%
  \BibitemOpen
  \bibfield  {author} {\bibinfo {author} {\bibfnamefont {Takol}\ \bibnamefont
  {Tangphati}}, \bibinfo {author} {\bibfnamefont {Indrani}\ \bibnamefont
  {Karar}}, \bibinfo {author} {\bibfnamefont {Anirudh}\ \bibnamefont
  {Pradhan}}, \ and\ \bibinfo {author} {\bibfnamefont {Ayan}\ \bibnamefont
  {Banerjee}},\ }\bibfield  {title} {\enquote {\bibinfo {title} {Constraints on
  the maximum mass of quark star and the {{GW}} 190814 event},}\ }\href
  {\doibase 10.1140/epjc/s10052-022-10024-6} {\bibfield  {journal} {\bibinfo
  {journal} {Eur. Phys. J. C}\ }\textbf {\bibinfo {volume} {82}},\ \bibinfo
  {pages} {57} (\bibinfo {year} {2022}{\natexlab{a}})}\BibitemShut {NoStop}%
\bibitem [{\citenamefont {Carvalho}\ \emph {et~al.}(2022)\citenamefont
  {Carvalho}, \citenamefont {Lobato}, \citenamefont {Moraes}, \citenamefont
  {Deb},\ and\ \citenamefont {Malheiro}}]{Carvalho:2022kxq}%
  \BibitemOpen
  \bibfield  {author} {\bibinfo {author} {\bibfnamefont {G.~A.}\ \bibnamefont
  {Carvalho}}, \bibinfo {author} {\bibfnamefont {R.~V.}\ \bibnamefont
  {Lobato}}, \bibinfo {author} {\bibfnamefont {P.~H. R.~S.}\ \bibnamefont
  {Moraes}}, \bibinfo {author} {\bibfnamefont {D.}~\bibnamefont {Deb}}, \ and\
  \bibinfo {author} {\bibfnamefont {M.}~\bibnamefont {Malheiro}},\ }\bibfield
  {title} {\enquote {\bibinfo {title} {Quark stars with 2.6
  \${{M}}\_{\textbackslash}{\o}dot\$ in a non-minimal geometry-matter coupling
  theory of gravity},}\ }\href {\doibase 10.1140/epjc/s10052-022-11058-6}
  {\bibfield  {journal} {\bibinfo  {journal} {Eur. Phys. J. C}\ }\textbf
  {\bibinfo {volume} {82}},\ \bibinfo {pages} {1096} (\bibinfo {year}
  {2022})}\BibitemShut {NoStop}%
\bibitem [{\citenamefont {Drischler}\ \emph
  {et~al.}(2021{\natexlab{a}})\citenamefont {Drischler}, \citenamefont {Holt},\
  and\ \citenamefont {Wellenhofer}}]{Drischler:2021kxf}%
  \BibitemOpen
  \bibfield  {author} {\bibinfo {author} {\bibfnamefont {C.}~\bibnamefont
  {Drischler}}, \bibinfo {author} {\bibfnamefont {J.~W.}\ \bibnamefont {Holt}},
  \ and\ \bibinfo {author} {\bibfnamefont {C.}~\bibnamefont {Wellenhofer}},\
  }\bibfield  {title} {\enquote {\bibinfo {title} {Chiral effective field
  theory and the high-density nuclear equation of state},}\ }\href {\doibase
  10.1146/annurev-nucl-102419-041903} {\bibfield  {journal} {\bibinfo
  {journal} {Ann. Rev. Nucl. Part. Sci.}\ }\textbf {\bibinfo {volume} {71}},\
  \bibinfo {pages} {403--432} (\bibinfo {year} {2021}{\natexlab{a}})},\ \Eprint
  {http://arxiv.org/abs/2101.01709} {arXiv:2101.01709 [nucl-th]} \BibitemShut
  {NoStop}%
\bibitem [{\citenamefont {Pretel}\ \emph {et~al.}(2025)\citenamefont {Pretel},
  \citenamefont {Tangphati}, \citenamefont {Sakall{\i}},\ and\ \citenamefont
  {Banerjee}}]{Pretel:2025roz}%
  \BibitemOpen
  \bibfield  {author} {\bibinfo {author} {\bibfnamefont {Juan M.~Z.}\
  \bibnamefont {Pretel}}, \bibinfo {author} {\bibfnamefont {Takol}\
  \bibnamefont {Tangphati}}, \bibinfo {author} {\bibfnamefont {{\.I}zzet}\
  \bibnamefont {Sakall{\i}}}, \ and\ \bibinfo {author} {\bibfnamefont {Ayan}\
  \bibnamefont {Banerjee}},\ }\bibfield  {title} {\enquote {\bibinfo {title}
  {White dwarfs in regularized {{4D Einstein-Gauss-Bonnet}} gravity},}\ }\href
  {\doibase 10.1016/j.physletb.2025.139581} {\bibfield  {journal} {\bibinfo
  {journal} {Phys. Lett. B}\ }\textbf {\bibinfo {volume} {866}},\ \bibinfo
  {pages} {139581} (\bibinfo {year} {2025})}\BibitemShut {NoStop}%
\bibitem [{\citenamefont {Pretel}\ \emph {et~al.}(2024)\citenamefont {Pretel},
  \citenamefont {Tangphati}, \citenamefont {Banerjee},\ and\ \citenamefont
  {Pradhan}}]{Pretel:2023nlr}%
  \BibitemOpen
  \bibfield  {author} {\bibinfo {author} {\bibfnamefont {Juan M.~Z.}\
  \bibnamefont {Pretel}}, \bibinfo {author} {\bibfnamefont {Takol}\
  \bibnamefont {Tangphati}}, \bibinfo {author} {\bibfnamefont {Ayan}\
  \bibnamefont {Banerjee}}, \ and\ \bibinfo {author} {\bibfnamefont {Anirudh}\
  \bibnamefont {Pradhan}},\ }\bibfield  {title} {\enquote {\bibinfo {title}
  {Effects of anisotropic pressure on interacting quark star structure},}\
  }\href {\doibase 10.1016/j.physletb.2023.138375} {\bibfield  {journal}
  {\bibinfo  {journal} {Phys. Lett. B}\ }\textbf {\bibinfo {volume} {848}},\
  \bibinfo {pages} {138375} (\bibinfo {year} {2024})}\BibitemShut {NoStop}%
\bibitem [{\citenamefont {Tangphati}\ \emph {et~al.}(2023)\citenamefont
  {Tangphati}, \citenamefont {Panotopoulos}, \citenamefont {Banerjee},\ and\
  \citenamefont {Pradhan}}]{Tangphati:2023efy}%
  \BibitemOpen
  \bibfield  {author} {\bibinfo {author} {\bibfnamefont {Takol}\ \bibnamefont
  {Tangphati}}, \bibinfo {author} {\bibfnamefont {Grigoris}\ \bibnamefont
  {Panotopoulos}}, \bibinfo {author} {\bibfnamefont {Ayan}\ \bibnamefont
  {Banerjee}}, \ and\ \bibinfo {author} {\bibfnamefont {Anirudh}\ \bibnamefont
  {Pradhan}},\ }\bibfield  {title} {\enquote {\bibinfo {title} {Charged compact
  stars with color--flavor-locked strange quark matter in f({{R}},{{T}})
  gravity},}\ }\href {\doibase 10.1016/j.cjph.2022.12.014} {\bibfield
  {journal} {\bibinfo  {journal} {Chin. J. Phys.}\ }\textbf {\bibinfo {volume}
  {82}},\ \bibinfo {pages} {62--74} (\bibinfo {year} {2023})}\BibitemShut
  {NoStop}%
\bibitem [{\citenamefont {Tangphati}\ \emph
  {et~al.}(2022{\natexlab{b}})\citenamefont {Tangphati}, \citenamefont {Karar},
  \citenamefont {Banerjee},\ and\ \citenamefont {Pradhan}}]{Tangphati:2022acb}%
  \BibitemOpen
  \bibfield  {author} {\bibinfo {author} {\bibfnamefont {Takol}\ \bibnamefont
  {Tangphati}}, \bibinfo {author} {\bibfnamefont {Indrani}\ \bibnamefont
  {Karar}}, \bibinfo {author} {\bibfnamefont {Ayan}\ \bibnamefont {Banerjee}},
  \ and\ \bibinfo {author} {\bibfnamefont {Anirudh}\ \bibnamefont {Pradhan}},\
  }\bibfield  {title} {\enquote {\bibinfo {title} {The mass--radius relation
  for quark stars in energy--momentum squared gravity},}\ }\href {\doibase
  10.1016/j.aop.2022.169149} {\bibfield  {journal} {\bibinfo  {journal} {Annals
  Phys.}\ }\textbf {\bibinfo {volume} {447}},\ \bibinfo {pages} {169149}
  (\bibinfo {year} {2022}{\natexlab{b}})}\BibitemShut {NoStop}%
\bibitem [{\citenamefont {Yuan}\ \emph {et~al.}(2022)\citenamefont {Yuan},
  \citenamefont {Li}, \citenamefont {Miao}, \citenamefont {Zuo},\ and\
  \citenamefont {Bai}}]{Yuan:2022dxb}%
  \BibitemOpen
  \bibfield  {author} {\bibinfo {author} {\bibfnamefont {Wen-Li}\ \bibnamefont
  {Yuan}}, \bibinfo {author} {\bibfnamefont {Ang}\ \bibnamefont {Li}}, \bibinfo
  {author} {\bibfnamefont {Zhiqiang}\ \bibnamefont {Miao}}, \bibinfo {author}
  {\bibfnamefont {Bingjun}\ \bibnamefont {Zuo}}, \ and\ \bibinfo {author}
  {\bibfnamefont {Zhan}\ \bibnamefont {Bai}},\ }\bibfield  {title} {\enquote
  {\bibinfo {title} {Interacting ud and uds quark matter at finite densities
  and quark stars},}\ }\href {\doibase 10.1103/PhysRevD.105.123004} {\bibfield
  {journal} {\bibinfo  {journal} {Phys. Rev. D}\ }\textbf {\bibinfo {volume}
  {105}},\ \bibinfo {pages} {123004} (\bibinfo {year} {2022})},\ \Eprint
  {http://arxiv.org/abs/2203.04798} {arXiv:2203.04798 [nucl-th]} \BibitemShut
  {NoStop}%
\bibitem [{\citenamefont {Xia}\ \emph {et~al.}(2022)\citenamefont {Xia},
  \citenamefont {Maruyama}, \citenamefont {Li}, \citenamefont {Sun},
  \citenamefont {Long},\ and\ \citenamefont {Zhang}}]{Xia:2022dvw}%
  \BibitemOpen
  \bibfield  {author} {\bibinfo {author} {\bibfnamefont {Cheng-Jun}\
  \bibnamefont {Xia}}, \bibinfo {author} {\bibfnamefont {Toshiki}\ \bibnamefont
  {Maruyama}}, \bibinfo {author} {\bibfnamefont {Ang}\ \bibnamefont {Li}},
  \bibinfo {author} {\bibfnamefont {Bao~Yuan}\ \bibnamefont {Sun}}, \bibinfo
  {author} {\bibfnamefont {Wen-Hui}\ \bibnamefont {Long}}, \ and\ \bibinfo
  {author} {\bibfnamefont {Ying-Xun}\ \bibnamefont {Zhang}},\ }\bibfield
  {title} {\enquote {\bibinfo {title} {Unified neutron star {{EOSs}} and
  neutron star structures in {{RMF}} models},}\ }\href {\doibase
  10.1088/1572-9494/ac71fd} {\bibfield  {journal} {\bibinfo  {journal} {Commun.
  Theor. Phys.}\ }\textbf {\bibinfo {volume} {74}},\ \bibinfo {pages} {095303}
  (\bibinfo {year} {2022})}\BibitemShut {NoStop}%
\bibitem [{\citenamefont {Chu}\ \emph {et~al.}(2025)\citenamefont {Chu},
  \citenamefont {Liu}, \citenamefont {Liu}, \citenamefont {Ju}, \citenamefont
  {Wu}, \citenamefont {Zhou},\ and\ \citenamefont {Li}}]{Chu:2025tsx}%
  \BibitemOpen
  \bibfield  {author} {\bibinfo {author} {\bibfnamefont {Peng-Cheng}\
  \bibnamefont {Chu}}, \bibinfo {author} {\bibfnamefont {He}~\bibnamefont
  {Liu}}, \bibinfo {author} {\bibfnamefont {Hong-Ming}\ \bibnamefont {Liu}},
  \bibinfo {author} {\bibfnamefont {Min}\ \bibnamefont {Ju}}, \bibinfo {author}
  {\bibfnamefont {Xu-Hao}\ \bibnamefont {Wu}}, \bibinfo {author} {\bibfnamefont
  {Ying}\ \bibnamefont {Zhou}}, \ and\ \bibinfo {author} {\bibfnamefont
  {Xiao-Hua}\ \bibnamefont {Li}},\ }\bibfield  {title} {\enquote {\bibinfo
  {title} {Quark star matter within an extended quasiparticle model},}\ }\href
  {\doibase 10.1140/epjc/s10052-025-14193-y} {\bibfield  {journal} {\bibinfo
  {journal} {Eur. Phys. J. C}\ }\textbf {\bibinfo {volume} {85}},\ \bibinfo
  {pages} {466} (\bibinfo {year} {2025})}\BibitemShut {NoStop}%
\bibitem [{\citenamefont {Liu}\ \emph {et~al.}(2024)\citenamefont {Liu},
  \citenamefont {Yang}, \citenamefont {Yuan}, \citenamefont {Ju}, \citenamefont
  {Wu},\ and\ \citenamefont {Chu}}]{Liu:2023ocg}%
  \BibitemOpen
  \bibfield  {author} {\bibinfo {author} {\bibfnamefont {He}~\bibnamefont
  {Liu}}, \bibinfo {author} {\bibfnamefont {Yong-Hang}\ \bibnamefont {Yang}},
  \bibinfo {author} {\bibfnamefont {Chi}\ \bibnamefont {Yuan}}, \bibinfo
  {author} {\bibfnamefont {Min}\ \bibnamefont {Ju}}, \bibinfo {author}
  {\bibfnamefont {Xu-Hao}\ \bibnamefont {Wu}}, \ and\ \bibinfo {author}
  {\bibfnamefont {Peng-Cheng}\ \bibnamefont {Chu}},\ }\bibfield  {title}
  {\enquote {\bibinfo {title} {Speed of sound and polytropic index in the
  {{Polyakov-loop Nambu-Jona-Lasinio}} model},}\ }\href {\doibase
  10.1103/PhysRevD.109.074037} {\bibfield  {journal} {\bibinfo  {journal}
  {Phys. Rev. D}\ }\textbf {\bibinfo {volume} {109}},\ \bibinfo {pages}
  {074037} (\bibinfo {year} {2024})},\ \Eprint
  {http://arxiv.org/abs/2311.10059} {arXiv:2311.10059 [hep-ph]} \BibitemShut
  {NoStop}%
\bibitem [{\citenamefont {Alford}\ \emph {et~al.}(2008)\citenamefont {Alford},
  \citenamefont {Schmitt}, \citenamefont {Rajagopal},\ and\ \citenamefont
  {Sch{\"a}fer}}]{Alford:2007xm}%
  \BibitemOpen
  \bibfield  {author} {\bibinfo {author} {\bibfnamefont {Mark~G.}\ \bibnamefont
  {Alford}}, \bibinfo {author} {\bibfnamefont {Andreas}\ \bibnamefont
  {Schmitt}}, \bibinfo {author} {\bibfnamefont {Krishna}\ \bibnamefont
  {Rajagopal}}, \ and\ \bibinfo {author} {\bibfnamefont {Thomas}\ \bibnamefont
  {Sch{\"a}fer}},\ }\bibfield  {title} {\enquote {\bibinfo {title} {Color
  superconductivity in dense quark matter},}\ }\href {\doibase
  10.1103/RevModPhys.80.1455} {\bibfield  {journal} {\bibinfo  {journal} {Rev.
  Mod. Phys.}\ }\textbf {\bibinfo {volume} {80}},\ \bibinfo {pages}
  {1455--1515} (\bibinfo {year} {2008})},\ \Eprint
  {http://arxiv.org/abs/0709.4635} {arXiv:0709.4635 [hep-ph]} \BibitemShut
  {NoStop}%
\bibitem [{\citenamefont {Fukushima}\ and\ \citenamefont
  {Hatsuda}(2011)}]{Fukushima:2010bq}%
  \BibitemOpen
  \bibfield  {author} {\bibinfo {author} {\bibfnamefont {Kenji}\ \bibnamefont
  {Fukushima}}\ and\ \bibinfo {author} {\bibfnamefont {Tetsuo}\ \bibnamefont
  {Hatsuda}},\ }\bibfield  {title} {\enquote {\bibinfo {title} {The phase
  diagram of dense {{QCD}}},}\ }\href {\doibase 10.1088/0034-4885/74/1/014001}
  {\bibfield  {journal} {\bibinfo  {journal} {Rept. Prog. Phys.}\ }\textbf
  {\bibinfo {volume} {74}},\ \bibinfo {pages} {014001} (\bibinfo {year}
  {2011})},\ \Eprint {http://arxiv.org/abs/1005.4814} {arXiv:1005.4814
  [hep-ph]} \BibitemShut {NoStop}%
\bibitem [{\citenamefont {G{\"a}rtlein}\ \emph {et~al.}(2025)\citenamefont
  {G{\"a}rtlein}, \citenamefont {Ivanytskyi}, \citenamefont {Sagun},\ and\
  \citenamefont {Lopes}}]{gartlein2025Colorsuperconducting}%
  \BibitemOpen
  \bibfield  {author} {\bibinfo {author} {\bibfnamefont {Christoph}\
  \bibnamefont {G{\"a}rtlein}}, \bibinfo {author} {\bibfnamefont {Oleksii}\
  \bibnamefont {Ivanytskyi}}, \bibinfo {author} {\bibfnamefont {Violetta}\
  \bibnamefont {Sagun}}, \ and\ \bibinfo {author} {\bibfnamefont {Il{\'i}dio}\
  \bibnamefont {Lopes}},\ }\bibfield  {title} {\enquote {\bibinfo {title}
  {Color-superconducting quarkyonic matter},}\ }\href@noop {} {\  (\bibinfo
  {year} {2025})}\BibitemShut {NoStop}%
\bibitem [{\citenamefont {Ivanytskyi}\ and\ \citenamefont
  {Blaschke}(2022)}]{Ivanytskyi:2022oxv}%
  \BibitemOpen
  \bibfield  {author} {\bibinfo {author} {\bibfnamefont {Oleksii}\ \bibnamefont
  {Ivanytskyi}}\ and\ \bibinfo {author} {\bibfnamefont {David}\ \bibnamefont
  {Blaschke}},\ }\bibfield  {title} {\enquote {\bibinfo {title} {Density
  functional approach to quark matter with confinement and color
  superconductivity},}\ }\href {\doibase 10.1103/PhysRevD.105.114042}
  {\bibfield  {journal} {\bibinfo  {journal} {Phys. Rev. D}\ }\textbf {\bibinfo
  {volume} {105}},\ \bibinfo {pages} {114042} (\bibinfo {year}
  {2022})}\BibitemShut {NoStop}%
\bibitem [{\citenamefont {Wen}\ \emph {et~al.}(2005)\citenamefont {Wen},
  \citenamefont {Zhong}, \citenamefont {Peng}, \citenamefont {Shen},\ and\
  \citenamefont {Ning}}]{Wen:2005uf}%
  \BibitemOpen
  \bibfield  {author} {\bibinfo {author} {\bibfnamefont {X.~J.}\ \bibnamefont
  {Wen}}, \bibinfo {author} {\bibfnamefont {X.~H.}\ \bibnamefont {Zhong}},
  \bibinfo {author} {\bibfnamefont {G.~X.}\ \bibnamefont {Peng}}, \bibinfo
  {author} {\bibfnamefont {P.~N.}\ \bibnamefont {Shen}}, \ and\ \bibinfo
  {author} {\bibfnamefont {P.~Z.}\ \bibnamefont {Ning}},\ }\bibfield  {title}
  {\enquote {\bibinfo {title} {Thermodynamics with density and temperature
  dependent particle masses and properties of bulk strange quark matter and
  strangelets},}\ }\href {\doibase 10.1103/PhysRevC.72.015204} {\bibfield
  {journal} {\bibinfo  {journal} {Phys. Rev. C}\ }\textbf {\bibinfo {volume}
  {72}},\ \bibinfo {pages} {015204} (\bibinfo {year} {2005})},\ \Eprint
  {http://arxiv.org/abs/hep-ph/0506050} {arXiv:hep-ph/0506050} \BibitemShut
  {NoStop}%
\bibitem [{\citenamefont {Chu}\ and\ \citenamefont {Chen}(2014)}]{Chu:2012rd}%
  \BibitemOpen
  \bibfield  {author} {\bibinfo {author} {\bibfnamefont {Peng-Cheng}\
  \bibnamefont {Chu}}\ and\ \bibinfo {author} {\bibfnamefont {Lie-Wen}\
  \bibnamefont {Chen}},\ }\bibfield  {title} {\enquote {\bibinfo {title} {Quark
  matter symmetry energy and quark stars},}\ }\href {\doibase
  10.1088/0004-637X/780/2/135} {\bibfield  {journal} {\bibinfo  {journal}
  {Astrophys. J.}\ }\textbf {\bibinfo {volume} {780}},\ \bibinfo {pages} {135}
  (\bibinfo {year} {2014})},\ \Eprint {http://arxiv.org/abs/1212.1388}
  {arXiv:1212.1388 [astro-ph.SR]} \BibitemShut {NoStop}%
\bibitem [{\citenamefont {Lu}\ \emph {et~al.}(2016)\citenamefont {Lu},
  \citenamefont {Peng}, \citenamefont {Xu},\ and\ \citenamefont
  {Zhang}}]{Lu:2016fki}%
  \BibitemOpen
  \bibfield  {author} {\bibinfo {author} {\bibfnamefont {Zhen-Yan}\
  \bibnamefont {Lu}}, \bibinfo {author} {\bibfnamefont {Guang-Xiong}\
  \bibnamefont {Peng}}, \bibinfo {author} {\bibfnamefont {Jian-Feng}\
  \bibnamefont {Xu}}, \ and\ \bibinfo {author} {\bibfnamefont {Shi-Peng}\
  \bibnamefont {Zhang}},\ }\bibfield  {title} {\enquote {\bibinfo {title}
  {Properties of quark matter in a new quasiparticle model with {{QCD}} running
  coupling},}\ }\href {\doibase 10.1007/s11433-015-0524-2} {\bibfield
  {journal} {\bibinfo  {journal} {Sci. China Phys. Mech. Astron.}\ }\textbf
  {\bibinfo {volume} {59}},\ \bibinfo {pages} {662001} (\bibinfo {year}
  {2016})}\BibitemShut {NoStop}%
\bibitem [{\citenamefont {Alba}\ \emph {et~al.}(2014)\citenamefont {Alba},
  \citenamefont {Alberico}, \citenamefont {Bluhm}, \citenamefont {Greco},
  \citenamefont {Ratti},\ and\ \citenamefont {Ruggieri}}]{Alba:2014lda}%
  \BibitemOpen
  \bibfield  {author} {\bibinfo {author} {\bibfnamefont {Paolo}\ \bibnamefont
  {Alba}}, \bibinfo {author} {\bibfnamefont {Wanda}\ \bibnamefont {Alberico}},
  \bibinfo {author} {\bibfnamefont {Marcus}\ \bibnamefont {Bluhm}}, \bibinfo
  {author} {\bibfnamefont {Vincenzo}\ \bibnamefont {Greco}}, \bibinfo {author}
  {\bibfnamefont {Claudia}\ \bibnamefont {Ratti}}, \ and\ \bibinfo {author}
  {\bibfnamefont {Marco}\ \bibnamefont {Ruggieri}},\ }\bibfield  {title}
  {\enquote {\bibinfo {title} {Polyakov loop and gluon quasiparticles: {{A}}
  self-consistent approach to {{Yang-Mills}} thermodynamics},}\ }\href
  {\doibase 10.1016/j.nuclphysa.2014.11.011} {\bibfield  {journal} {\bibinfo
  {journal} {Nucl. Phys. A}\ }\textbf {\bibinfo {volume} {934}},\ \bibinfo
  {pages} {41--51} (\bibinfo {year} {2014})},\ \Eprint
  {http://arxiv.org/abs/1402.6213} {arXiv:1402.6213 [hep-ph]} \BibitemShut
  {NoStop}%
\bibitem [{\citenamefont {Pal}\ and\ \citenamefont
  {Chaudhuri}(2023)}]{Pal:2023dlv}%
  \BibitemOpen
  \bibfield  {author} {\bibinfo {author} {\bibfnamefont {Suman}\ \bibnamefont
  {Pal}}\ and\ \bibinfo {author} {\bibfnamefont {Gargi}\ \bibnamefont
  {Chaudhuri}},\ }\bibfield  {title} {\enquote {\bibinfo {title} {Medium
  effects in the {{MIT}} bag model for quark matter: {{Self-consistent}}
  thermodynamical treatment},}\ }\href {\doibase 10.1103/PhysRevD.108.103028}
  {\bibfield  {journal} {\bibinfo  {journal} {Phys. Rev. D}\ }\textbf {\bibinfo
  {volume} {108}},\ \bibinfo {pages} {103028} (\bibinfo {year}
  {2023})}\BibitemShut {NoStop}%
\bibitem [{\citenamefont {Xu}\ \emph {et~al.}(2015{\natexlab{a}})\citenamefont
  {Xu}, \citenamefont {Peng}, \citenamefont {Liu}, \citenamefont {Hou},\ and\
  \citenamefont {Chen}}]{Xu:2015wya}%
  \BibitemOpen
  \bibfield  {author} {\bibinfo {author} {\bibfnamefont {J.~F.}\ \bibnamefont
  {Xu}}, \bibinfo {author} {\bibfnamefont {G.~X.}\ \bibnamefont {Peng}},
  \bibinfo {author} {\bibfnamefont {F.}~\bibnamefont {Liu}}, \bibinfo {author}
  {\bibfnamefont {De-Fu}\ \bibnamefont {Hou}}, \ and\ \bibinfo {author}
  {\bibfnamefont {Lie-Wen}\ \bibnamefont {Chen}},\ }\bibfield  {title}
  {\enquote {\bibinfo {title} {Strange matter and strange stars in a
  thermodynamically self-consistent perturbation model with running coupling
  and running strange quark mass},}\ }\href {\doibase
  10.1103/PhysRevD.92.025025} {\bibfield  {journal} {\bibinfo  {journal} {Phys.
  Rev. D}\ }\textbf {\bibinfo {volume} {92}},\ \bibinfo {pages} {025025}
  (\bibinfo {year} {2015}{\natexlab{a}})},\ \Eprint
  {http://arxiv.org/abs/1512.08229} {arXiv:1512.08229 [hep-ph]} \BibitemShut
  {NoStop}%
\bibitem [{\citenamefont {Carignano}\ \emph {et~al.}(2016)\citenamefont
  {Carignano}, \citenamefont {Mammarella},\ and\ \citenamefont
  {Mannarelli}}]{Carignano:2016rvs}%
  \BibitemOpen
  \bibfield  {author} {\bibinfo {author} {\bibfnamefont {Stefano}\ \bibnamefont
  {Carignano}}, \bibinfo {author} {\bibfnamefont {Andrea}\ \bibnamefont
  {Mammarella}}, \ and\ \bibinfo {author} {\bibfnamefont {Massimo}\
  \bibnamefont {Mannarelli}},\ }\bibfield  {title} {\enquote {\bibinfo {title}
  {Equation of state of imbalanced cold matter from chiral perturbation
  theory},}\ }\href {\doibase 10.1103/PhysRevD.93.051503} {\bibfield  {journal}
  {\bibinfo  {journal} {Phys. Rev. D}\ }\textbf {\bibinfo {volume} {93}},\
  \bibinfo {pages} {051503} (\bibinfo {year} {2016})},\ \Eprint
  {http://arxiv.org/abs/1602.01317} {arXiv:1602.01317 [hep-ph]} \BibitemShut
  {NoStop}%
\bibitem [{\citenamefont {Graf}\ \emph {et~al.}(2016)\citenamefont {Graf},
  \citenamefont {{Schaffner-Bielich}},\ and\ \citenamefont
  {Fraga}}]{Graf:2015pyl}%
  \BibitemOpen
  \bibfield  {author} {\bibinfo {author} {\bibfnamefont {Thorben}\ \bibnamefont
  {Graf}}, \bibinfo {author} {\bibfnamefont {Juergen}\ \bibnamefont
  {{Schaffner-Bielich}}}, \ and\ \bibinfo {author} {\bibfnamefont {Eduardo~S.}\
  \bibnamefont {Fraga}},\ }\bibfield  {title} {\enquote {\bibinfo {title}
  {Perturbative thermodynamics at nonzero isospin density for cold {{QCD}}},}\
  }\href {\doibase 10.1103/PhysRevD.93.085030} {\bibfield  {journal} {\bibinfo
  {journal} {Phys. Rev. D}\ }\textbf {\bibinfo {volume} {93}},\ \bibinfo
  {pages} {085030} (\bibinfo {year} {2016})},\ \Eprint
  {http://arxiv.org/abs/1511.09457} {arXiv:1511.09457 [hep-ph]} \BibitemShut
  {NoStop}%
\bibitem [{\citenamefont {Peng}(2005)}]{Peng:2005xp}%
  \BibitemOpen
  \bibfield  {author} {\bibinfo {author} {\bibfnamefont {G.~X.}\ \bibnamefont
  {Peng}},\ }\bibfield  {title} {\enquote {\bibinfo {title} {Thermodynamic
  correction to the strong interaction in the perturbative regime},}\ }\href
  {\doibase 10.1209/epl/i2005-10189-8} {\bibfield  {journal} {\bibinfo
  {journal} {Europhys. Lett.}\ }\textbf {\bibinfo {volume} {72}},\ \bibinfo
  {pages} {69--75} (\bibinfo {year} {2005})}\BibitemShut {NoStop}%
\bibitem [{\citenamefont {Xu}\ \emph {et~al.}(2015{\natexlab{b}})\citenamefont
  {Xu}, \citenamefont {Peng}, \citenamefont {Lu},\ and\ \citenamefont
  {Cui}}]{Xu:2014zea}%
  \BibitemOpen
  \bibfield  {author} {\bibinfo {author} {\bibfnamefont {Jian-Feng}\
  \bibnamefont {Xu}}, \bibinfo {author} {\bibfnamefont {Guang-Xiong}\
  \bibnamefont {Peng}}, \bibinfo {author} {\bibfnamefont {Zhen-Yan}\
  \bibnamefont {Lu}}, \ and\ \bibinfo {author} {\bibfnamefont {Shuai-Shuai}\
  \bibnamefont {Cui}},\ }\bibfield  {title} {\enquote {\bibinfo {title}
  {Two-flavor quark matter in the perturbation theory with full thermodynamic
  consistency},}\ }\href {\doibase 10.1007/s11433-014-5599-6} {\bibfield
  {journal} {\bibinfo  {journal} {Sci. China Phys. Mech. Astron.}\ }\textbf
  {\bibinfo {volume} {58}},\ \bibinfo {pages} {042001} (\bibinfo {year}
  {2015}{\natexlab{b}})}\BibitemShut {NoStop}%
\bibitem [{\citenamefont {Fraga}\ \emph {et~al.}(2001)\citenamefont {Fraga},
  \citenamefont {Pisarski},\ and\ \citenamefont
  {{Schaffner-Bielich}}}]{Fraga:2001id}%
  \BibitemOpen
  \bibfield  {author} {\bibinfo {author} {\bibfnamefont {Eduardo~S.}\
  \bibnamefont {Fraga}}, \bibinfo {author} {\bibfnamefont {Robert~D.}\
  \bibnamefont {Pisarski}}, \ and\ \bibinfo {author} {\bibfnamefont {Jurgen}\
  \bibnamefont {{Schaffner-Bielich}}},\ }\bibfield  {title} {\enquote {\bibinfo
  {title} {Small, dense quark stars from perturbative {{QCD}}},}\ }\href
  {\doibase 10.1103/PhysRevD.63.121702} {\bibfield  {journal} {\bibinfo
  {journal} {Phys. Rev. D}\ }\textbf {\bibinfo {volume} {63}},\ \bibinfo
  {pages} {121702} (\bibinfo {year} {2001})},\ \Eprint
  {http://arxiv.org/abs/hep-ph/0101143} {arXiv:hep-ph/0101143} \BibitemShut
  {NoStop}%
\bibitem [{\citenamefont {Fraga}\ \emph {et~al.}(2002)\citenamefont {Fraga},
  \citenamefont {Pisarski},\ and\ \citenamefont
  {{Schaffner-Bielich}}}]{Fraga:2001xc}%
  \BibitemOpen
  \bibfield  {author} {\bibinfo {author} {\bibfnamefont {Eduardo~S.}\
  \bibnamefont {Fraga}}, \bibinfo {author} {\bibfnamefont {Robert~D.}\
  \bibnamefont {Pisarski}}, \ and\ \bibinfo {author} {\bibfnamefont {Jurgen}\
  \bibnamefont {{Schaffner-Bielich}}},\ }\bibfield  {title} {\enquote {\bibinfo
  {title} {New class of compact stars at high density},}\ }in\ \href {\doibase
  10.1016/S0375-9474(02)00709-1} {\emph {\bibinfo {booktitle} {Nucl. {{Phys}}.
  {{A}}}}},\ \bibinfo {series} {Nuclear {{Physics A}}}, Vol.\ \bibinfo {volume}
  {702}\ (\bibinfo {year} {2002})\ pp.\ \bibinfo {pages} {217--223},\ \Eprint
  {http://arxiv.org/abs/nucl-th/0110077} {arXiv:nucl-th/0110077} \BibitemShut
  {NoStop}%
\bibitem [{\citenamefont {Fraga}\ and\ \citenamefont
  {Romatschke}(2005)}]{Fraga:2004gz}%
  \BibitemOpen
  \bibfield  {author} {\bibinfo {author} {\bibfnamefont {Eduardo~S.}\
  \bibnamefont {Fraga}}\ and\ \bibinfo {author} {\bibfnamefont {Paul}\
  \bibnamefont {Romatschke}},\ }\bibfield  {title} {\enquote {\bibinfo {title}
  {The {{Role}} of quark mass in cold and dense perturbative {{QCD}}},}\ }\href
  {\doibase 10.1103/PhysRevD.71.105014} {\bibfield  {journal} {\bibinfo
  {journal} {Phys. Rev. D}\ }\textbf {\bibinfo {volume} {71}},\ \bibinfo
  {pages} {105014} (\bibinfo {year} {2005})},\ \Eprint
  {http://arxiv.org/abs/hep-ph/0412298} {arXiv:hep-ph/0412298} \BibitemShut
  {NoStop}%
\bibitem [{\citenamefont {Freedman}\ and\ \citenamefont
  {McLerran}(1977)}]{Freedman:1976ub}%
  \BibitemOpen
  \bibfield  {author} {\bibinfo {author} {\bibfnamefont {Barry~A.}\
  \bibnamefont {Freedman}}\ and\ \bibinfo {author} {\bibfnamefont {Larry~D.}\
  \bibnamefont {McLerran}},\ }\bibfield  {title} {\enquote {\bibinfo {title}
  {Fermions and {{Gauge Vector Mesons}} at {{Finite Temperature}} and
  {{Density}}. 3. {{The Ground State Energy}} of a {{Relativistic Quark
  Gas}}},}\ }\href {\doibase 10.1103/PhysRevD.16.1169} {\bibfield  {journal}
  {\bibinfo  {journal} {Phys. Rev. D}\ }\textbf {\bibinfo {volume} {16}},\
  \bibinfo {pages} {1169} (\bibinfo {year} {1977})}\BibitemShut {NoStop}%
\bibitem [{\citenamefont {Ma}\ \emph {et~al.}(2023)\citenamefont {Ma},
  \citenamefont {Lu}, \citenamefont {Xu}, \citenamefont {Peng}, \citenamefont
  {Fu},\ and\ \citenamefont {Wang}}]{Ma:2023stj}%
  \BibitemOpen
  \bibfield  {author} {\bibinfo {author} {\bibfnamefont {Zhi-Jun}\ \bibnamefont
  {Ma}}, \bibinfo {author} {\bibfnamefont {Zhen-Yan}\ \bibnamefont {Lu}},
  \bibinfo {author} {\bibfnamefont {Jian-Feng}\ \bibnamefont {Xu}}, \bibinfo
  {author} {\bibfnamefont {Guang-Xiong}\ \bibnamefont {Peng}}, \bibinfo
  {author} {\bibfnamefont {Xiangyun}\ \bibnamefont {Fu}}, \ and\ \bibinfo
  {author} {\bibfnamefont {Junnian}\ \bibnamefont {Wang}},\ }\bibfield  {title}
  {\enquote {\bibinfo {title} {Cold quark matter in a quasiparticle model:
  {{Thermodynamic}} consistency and stellar properties},}\ }\href {\doibase
  10.1103/PhysRevD.108.054017} {\bibfield  {journal} {\bibinfo  {journal}
  {Phys. Rev. D}\ }\textbf {\bibinfo {volume} {108}},\ \bibinfo {pages}
  {054017} (\bibinfo {year} {2023})},\ \Eprint
  {http://arxiv.org/abs/2308.05308} {arXiv:2308.05308 [hep-ph]} \BibitemShut
  {NoStop}%
\bibitem [{\citenamefont {Farhi}\ and\ \citenamefont
  {Jaffe}(1984)}]{Farhi:1984qu}%
  \BibitemOpen
  \bibfield  {author} {\bibinfo {author} {\bibfnamefont {Edward}\ \bibnamefont
  {Farhi}}\ and\ \bibinfo {author} {\bibfnamefont {R.~L.}\ \bibnamefont
  {Jaffe}},\ }\bibfield  {title} {\enquote {\bibinfo {title} {Strange
  matter},}\ }\href {\doibase 10.1103/PhysRevD.30.2379} {\bibfield  {journal}
  {\bibinfo  {journal} {Phys. Rev. D}\ }\textbf {\bibinfo {volume} {30}},\
  \bibinfo {pages} {2379} (\bibinfo {year} {1984})}\BibitemShut {NoStop}%
\bibitem [{\citenamefont {Lopes}\ \emph {et~al.}(2021)\citenamefont {Lopes},
  \citenamefont {Biesdorf},\ and\ \citenamefont {Menezes}}]{Lopes:2020btp}%
  \BibitemOpen
  \bibfield  {author} {\bibinfo {author} {\bibfnamefont {Luiz~L.}\ \bibnamefont
  {Lopes}}, \bibinfo {author} {\bibfnamefont {Carline}\ \bibnamefont
  {Biesdorf}}, \ and\ \bibinfo {author} {\bibfnamefont {D.~{\'e}bora~P.}\
  \bibnamefont {Menezes}},\ }\bibfield  {title} {\enquote {\bibinfo {title}
  {Modified {{MIT}} bag {{Models}}---part {{I}}: {{Thermodynamic}} consistency,
  stability windows and symmetry group},}\ }\href {\doibase
  10.1088/1402-4896/abef34} {\bibfield  {journal} {\bibinfo  {journal} {Phys.
  Scripta}\ }\textbf {\bibinfo {volume} {96}},\ \bibinfo {pages} {065303}
  (\bibinfo {year} {2021})},\ \Eprint {http://arxiv.org/abs/2005.13136}
  {arXiv:2005.13136 [hep-ph]} \BibitemShut {NoStop}%
\bibitem [{\citenamefont {Pal}\ \emph {et~al.}(2023)\citenamefont {Pal},
  \citenamefont {Podder}, \citenamefont {Sen},\ and\ \citenamefont
  {Chaudhuri}}]{Pal:2023quk}%
  \BibitemOpen
  \bibfield  {author} {\bibinfo {author} {\bibfnamefont {Suman}\ \bibnamefont
  {Pal}}, \bibinfo {author} {\bibfnamefont {Soumen}\ \bibnamefont {Podder}},
  \bibinfo {author} {\bibfnamefont {Debashree}\ \bibnamefont {Sen}}, \ and\
  \bibinfo {author} {\bibfnamefont {Gargi}\ \bibnamefont {Chaudhuri}},\
  }\bibfield  {title} {\enquote {\bibinfo {title} {Speed of sound in hybrid
  stars and the role of bag pressure in the emergence of special points on the
  {{M-R}} variation of hybrid stars},}\ }\href {\doibase
  10.1103/PhysRevD.107.063019} {\bibfield  {journal} {\bibinfo  {journal}
  {Phys. Rev. D}\ }\textbf {\bibinfo {volume} {107}},\ \bibinfo {pages}
  {063019} (\bibinfo {year} {2023})},\ \Eprint
  {http://arxiv.org/abs/2303.04653} {arXiv:2303.04653 [nucl-th]} \BibitemShut
  {NoStop}%
\bibitem [{\citenamefont {Hippert}\ \emph {et~al.}(2024)\citenamefont
  {Hippert}, \citenamefont {Grefa}, \citenamefont {Manning}, \citenamefont
  {Noronha}, \citenamefont {{Noronha-Hostler}}, \citenamefont
  {Portillo~Vazquez}, \citenamefont {Ratti}, \citenamefont {Rougemont},\ and\
  \citenamefont {Trujillo}}]{Hippert:2023bel}%
  \BibitemOpen
  \bibfield  {author} {\bibinfo {author} {\bibfnamefont {Mauricio}\
  \bibnamefont {Hippert}}, \bibinfo {author} {\bibfnamefont {Joaquin}\
  \bibnamefont {Grefa}}, \bibinfo {author} {\bibfnamefont {T.~Andrew}\
  \bibnamefont {Manning}}, \bibinfo {author} {\bibfnamefont {Jorge}\
  \bibnamefont {Noronha}}, \bibinfo {author} {\bibfnamefont {Jacquelyn}\
  \bibnamefont {{Noronha-Hostler}}}, \bibinfo {author} {\bibfnamefont {Israel}\
  \bibnamefont {Portillo~Vazquez}}, \bibinfo {author} {\bibfnamefont {Claudia}\
  \bibnamefont {Ratti}}, \bibinfo {author} {\bibfnamefont {Romulo}\
  \bibnamefont {Rougemont}}, \ and\ \bibinfo {author} {\bibfnamefont {Michael}\
  \bibnamefont {Trujillo}},\ }\bibfield  {title} {\enquote {\bibinfo {title}
  {Bayesian location of the {{QCD}} critical point from a holographic
  perspective},}\ }\href {\doibase 10.1103/PhysRevD.110.094006} {\bibfield
  {journal} {\bibinfo  {journal} {Phys. Rev. D}\ }\textbf {\bibinfo {volume}
  {110}},\ \bibinfo {pages} {094006} (\bibinfo {year} {2024})}\BibitemShut
  {NoStop}%
\bibitem [{\citenamefont {Marquez}\ \emph {et~al.}(2024)\citenamefont
  {Marquez}, \citenamefont {Malik}, \citenamefont {Pais}, \citenamefont
  {Menezes},\ and\ \citenamefont {Provid{\^e}ncia}}]{Marquez:2024bzj}%
  \BibitemOpen
  \bibfield  {author} {\bibinfo {author} {\bibfnamefont {K.~D.}\ \bibnamefont
  {Marquez}}, \bibinfo {author} {\bibfnamefont {Tuhin}\ \bibnamefont {Malik}},
  \bibinfo {author} {\bibfnamefont {Helena}\ \bibnamefont {Pais}}, \bibinfo
  {author} {\bibfnamefont {D{\'e}bora~P.}\ \bibnamefont {Menezes}}, \ and\
  \bibinfo {author} {\bibfnamefont {Constan{\c c}a}\ \bibnamefont
  {Provid{\^e}ncia}},\ }\bibfield  {title} {\enquote {\bibinfo {title}
  {Nambu--{{Jona-Lasinio}} description of hadronic matter from a {{Bayesian}}
  approach},}\ }\href {\doibase 10.1103/PhysRevD.110.063040} {\bibfield
  {journal} {\bibinfo  {journal} {Phys. Rev. D}\ }\textbf {\bibinfo {volume}
  {110}},\ \bibinfo {pages} {063040} (\bibinfo {year} {2024})}\BibitemShut
  {NoStop}%
\bibitem [{\citenamefont {Malik}\ and\ \citenamefont
  {Provid{\^e}ncia}(2022)}]{Malik:2022jqc}%
  \BibitemOpen
  \bibfield  {author} {\bibinfo {author} {\bibfnamefont {Tuhin}\ \bibnamefont
  {Malik}}\ and\ \bibinfo {author} {\bibfnamefont {Constan{\c c}a}\
  \bibnamefont {Provid{\^e}ncia}},\ }\bibfield  {title} {\enquote {\bibinfo
  {title} {Bayesian inference of signatures of hyperons inside neutron
  stars},}\ }\href {\doibase 10.1103/PhysRevD.106.063024} {\bibfield  {journal}
  {\bibinfo  {journal} {Phys. Rev. D}\ }\textbf {\bibinfo {volume} {106}},\
  \bibinfo {pages} {063024} (\bibinfo {year} {2022})}\BibitemShut {NoStop}%
\bibitem [{\citenamefont {Drischler}\ \emph {et~al.}(2024)\citenamefont
  {Drischler}, \citenamefont {Giuliani}, \citenamefont {Bezoui}, \citenamefont
  {Piekarewicz},\ and\ \citenamefont {Viens}}]{Drischler:2024ebw}%
  \BibitemOpen
  \bibfield  {author} {\bibinfo {author} {\bibfnamefont {C.}~\bibnamefont
  {Drischler}}, \bibinfo {author} {\bibfnamefont {P.~G.}\ \bibnamefont
  {Giuliani}}, \bibinfo {author} {\bibfnamefont {S.}~\bibnamefont {Bezoui}},
  \bibinfo {author} {\bibfnamefont {J.}~\bibnamefont {Piekarewicz}}, \ and\
  \bibinfo {author} {\bibfnamefont {F.}~\bibnamefont {Viens}},\ }\bibfield
  {title} {\enquote {\bibinfo {title} {Bayesian mixture model approach to
  quantifying the empirical nuclear saturation point},}\ }\href {\doibase
  10.1103/PhysRevC.110.044320} {\bibfield  {journal} {\bibinfo  {journal}
  {Phys. Rev. C}\ }\textbf {\bibinfo {volume} {110}},\ \bibinfo {pages}
  {044320} (\bibinfo {year} {2024})}\BibitemShut {NoStop}%
\bibitem [{\citenamefont {Semposki}\ \emph {et~al.}(2025)\citenamefont
  {Semposki}, \citenamefont {Drischler}, \citenamefont {Furnstahl},
  \citenamefont {Melendez},\ and\ \citenamefont {Phillips}}]{Semposki:2024vnp}%
  \BibitemOpen
  \bibfield  {author} {\bibinfo {author} {\bibfnamefont {A.~C.}\ \bibnamefont
  {Semposki}}, \bibinfo {author} {\bibfnamefont {C.}~\bibnamefont {Drischler}},
  \bibinfo {author} {\bibfnamefont {R.~J.}\ \bibnamefont {Furnstahl}}, \bibinfo
  {author} {\bibfnamefont {J.~A.}\ \bibnamefont {Melendez}}, \ and\ \bibinfo
  {author} {\bibfnamefont {D.~R.}\ \bibnamefont {Phillips}},\ }\bibfield
  {title} {\enquote {\bibinfo {title} {From chiral effective field theory to
  perturbative {{QCD}}: {{A Bayesian}} model mixing approach to symmetric
  nuclear matter},}\ }\href {\doibase 10.1103/PhysRevC.111.035804} {\bibfield
  {journal} {\bibinfo  {journal} {Phys. Rev. C}\ }\textbf {\bibinfo {volume}
  {111}},\ \bibinfo {pages} {035804} (\bibinfo {year} {2025})}\BibitemShut
  {NoStop}%
\bibitem [{\citenamefont {Drischler}\ \emph {et~al.}(2020)\citenamefont
  {Drischler}, \citenamefont {Furnstahl}, \citenamefont {Melendez},\ and\
  \citenamefont {Phillips}}]{Drischler:2020hwi}%
  \BibitemOpen
  \bibfield  {author} {\bibinfo {author} {\bibfnamefont {C.}~\bibnamefont
  {Drischler}}, \bibinfo {author} {\bibfnamefont {R.~J.}\ \bibnamefont
  {Furnstahl}}, \bibinfo {author} {\bibfnamefont {J.~A.}\ \bibnamefont
  {Melendez}}, \ and\ \bibinfo {author} {\bibfnamefont {D.~R.}\ \bibnamefont
  {Phillips}},\ }\bibfield  {title} {\enquote {\bibinfo {title} {How {{Well Do
  We Know}} the {{Neutron-Matter Equation}} of {{State}} at the {{Densities
  Inside Neutron Stars}}? {{A Bayesian Approach}} with {{Correlated
  Uncertainties}}},}\ }\href {\doibase 10.1103/PhysRevLett.125.202702}
  {\bibfield  {journal} {\bibinfo  {journal} {Phys. Rev. Lett.}\ }\textbf
  {\bibinfo {volume} {125}},\ \bibinfo {pages} {202702} (\bibinfo {year}
  {2020})}\BibitemShut {NoStop}%
\bibitem [{\citenamefont {Annala}\ \emph {et~al.}(2020)\citenamefont {Annala},
  \citenamefont {Gorda}, \citenamefont {Kurkela}, \citenamefont
  {N{\"a}ttil{\"a}},\ and\ \citenamefont {Vuorinen}}]{Annala:2019puf}%
  \BibitemOpen
  \bibfield  {author} {\bibinfo {author} {\bibfnamefont {Eemeli}\ \bibnamefont
  {Annala}}, \bibinfo {author} {\bibfnamefont {Tyler}\ \bibnamefont {Gorda}},
  \bibinfo {author} {\bibfnamefont {Aleksi}\ \bibnamefont {Kurkela}}, \bibinfo
  {author} {\bibfnamefont {Joonas}\ \bibnamefont {N{\"a}ttil{\"a}}}, \ and\
  \bibinfo {author} {\bibfnamefont {Aleksi}\ \bibnamefont {Vuorinen}},\
  }\bibfield  {title} {\enquote {\bibinfo {title} {Evidence for quark-matter
  cores in massive neutron stars},}\ }\href {\doibase
  10.1038/s41567-020-0914-9} {\bibfield  {journal} {\bibinfo  {journal} {Nature
  Phys.}\ }\textbf {\bibinfo {volume} {16}},\ \bibinfo {pages} {907--910}
  (\bibinfo {year} {2020})},\ \Eprint {http://arxiv.org/abs/1903.09121}
  {arXiv:1903.09121 [astro-ph.HE]} \BibitemShut {NoStop}%
\bibitem [{\citenamefont {Oppenheimer}\ and\ \citenamefont
  {Volkoff}(1939)}]{Oppenheimer:1939ne}%
  \BibitemOpen
  \bibfield  {author} {\bibinfo {author} {\bibfnamefont {J.~R.}\ \bibnamefont
  {Oppenheimer}}\ and\ \bibinfo {author} {\bibfnamefont {G.~M.}\ \bibnamefont
  {Volkoff}},\ }\bibfield  {title} {\enquote {\bibinfo {title} {On {{Massive}}
  neutron cores},}\ }\href {\doibase 10.1103/PhysRev.55.374} {\bibfield
  {journal} {\bibinfo  {journal} {Phys. Rev.}\ }\textbf {\bibinfo {volume}
  {55}},\ \bibinfo {pages} {374--381} (\bibinfo {year} {1939})}\BibitemShut
  {NoStop}%
\bibitem [{\citenamefont {Linares}\ \emph {et~al.}(2018)\citenamefont
  {Linares}, \citenamefont {Shahbaz},\ and\ \citenamefont
  {Casares}}]{Linares:2018ppq}%
  \BibitemOpen
  \bibfield  {author} {\bibinfo {author} {\bibfnamefont {Manuel}\ \bibnamefont
  {Linares}}, \bibinfo {author} {\bibfnamefont {Tariq}\ \bibnamefont
  {Shahbaz}}, \ and\ \bibinfo {author} {\bibfnamefont {Jorge}\ \bibnamefont
  {Casares}},\ }\bibfield  {title} {\enquote {\bibinfo {title} {Peering into
  the dark side: {{Magnesium}} lines establish a massive neutron star in {{PSR
  J2215}}+5135},}\ }\href {\doibase 10.3847/1538-4357/aabde6} {\bibfield
  {journal} {\bibinfo  {journal} {Astrophys. J.}\ }\textbf {\bibinfo {volume}
  {859}},\ \bibinfo {pages} {54} (\bibinfo {year} {2018})},\ \Eprint
  {http://arxiv.org/abs/1805.08799} {arXiv:1805.08799 [astro-ph.HE]}
  \BibitemShut {NoStop}%
\bibitem [{\citenamefont {N{\"a}ttil{\"a}}\ \emph {et~al.}(2017)\citenamefont
  {N{\"a}ttil{\"a}}, \citenamefont {Miller}, \citenamefont {Steiner},
  \citenamefont {Kajava}, \citenamefont {Suleimanov},\ and\ \citenamefont
  {Poutanen}}]{Nattila:2017wtj}%
  \BibitemOpen
  \bibfield  {author} {\bibinfo {author} {\bibfnamefont {J.}~\bibnamefont
  {N{\"a}ttil{\"a}}}, \bibinfo {author} {\bibfnamefont {M.~C.}\ \bibnamefont
  {Miller}}, \bibinfo {author} {\bibfnamefont {A.~W.}\ \bibnamefont {Steiner}},
  \bibinfo {author} {\bibfnamefont {J.~J.~E.}\ \bibnamefont {Kajava}}, \bibinfo
  {author} {\bibfnamefont {V.~F.}\ \bibnamefont {Suleimanov}}, \ and\ \bibinfo
  {author} {\bibfnamefont {J.}~\bibnamefont {Poutanen}},\ }\bibfield  {title}
  {\enquote {\bibinfo {title} {Neutron star mass and radius measurements from
  atmospheric model fits to {{X-ray}} burst cooling tail spectra},}\ }\href
  {\doibase 10.1051/0004-6361/201731082} {\bibfield  {journal} {\bibinfo
  {journal} {Astron. Astrophys.}\ }\textbf {\bibinfo {volume} {608}},\ \bibinfo
  {pages} {A31} (\bibinfo {year} {2017})}\BibitemShut {NoStop}%
\bibitem [{\citenamefont {Miller}\ \emph {et~al.}(2019)\citenamefont {Miller}
  \emph {et~al.}}]{Miller:2019cac}%
  \BibitemOpen
  \bibfield  {author} {\bibinfo {author} {\bibfnamefont {M.~C.}\ \bibnamefont
  {Miller}} \emph {et~al.},\ }\bibfield  {title} {\enquote {\bibinfo {title}
  {{{PSR J0030}}+0451 {{Mass}} and {{Radius}} from {{NICER Data}} and
  {{Implications}} for the {{Properties}} of {{Neutron Star Matter}}},}\ }\href
  {\doibase 10.3847/2041-8213/ab50c5} {\bibfield  {journal} {\bibinfo
  {journal} {Astrophys. J. Lett.}\ }\textbf {\bibinfo {volume} {887}},\
  \bibinfo {pages} {L24} (\bibinfo {year} {2019})}\BibitemShut {NoStop}%
\bibitem [{\citenamefont {Cao}\ \emph {et~al.}(2022)\citenamefont {Cao},
  \citenamefont {Chen}, \citenamefont {Chu},\ and\ \citenamefont
  {Zhou}}]{Cao:2020zxi}%
  \BibitemOpen
  \bibfield  {author} {\bibinfo {author} {\bibfnamefont {Zheng}\ \bibnamefont
  {Cao}}, \bibinfo {author} {\bibfnamefont {Lie-Wen}\ \bibnamefont {Chen}},
  \bibinfo {author} {\bibfnamefont {Peng-Cheng}\ \bibnamefont {Chu}}, \ and\
  \bibinfo {author} {\bibfnamefont {Ying}\ \bibnamefont {Zhou}},\ }\bibfield
  {title} {\enquote {\bibinfo {title} {{{GW190814}}: {{Circumstantial}}
  evidence for up-down quark star},}\ }\href {\doibase
  10.1103/PhysRevD.106.083007} {\bibfield  {journal} {\bibinfo  {journal}
  {Phys. Rev. D}\ }\textbf {\bibinfo {volume} {106}},\ \bibinfo {pages}
  {083007} (\bibinfo {year} {2022})},\ \Eprint
  {http://arxiv.org/abs/2009.00942} {arXiv:2009.00942 [astro-ph.HE]}
  \BibitemShut {NoStop}%
\bibitem [{\citenamefont {Drischler}\ \emph
  {et~al.}(2021{\natexlab{b}})\citenamefont {Drischler}, \citenamefont {Han},
  \citenamefont {Lattimer}, \citenamefont {Prakash}, \citenamefont {Reddy},\
  and\ \citenamefont {Zhao}}]{Drischler:2020fvz}%
  \BibitemOpen
  \bibfield  {author} {\bibinfo {author} {\bibfnamefont {Christian}\
  \bibnamefont {Drischler}}, \bibinfo {author} {\bibfnamefont {Sophia}\
  \bibnamefont {Han}}, \bibinfo {author} {\bibfnamefont {James~M.}\
  \bibnamefont {Lattimer}}, \bibinfo {author} {\bibfnamefont {Madappa}\
  \bibnamefont {Prakash}}, \bibinfo {author} {\bibfnamefont {Sanjay}\
  \bibnamefont {Reddy}}, \ and\ \bibinfo {author} {\bibfnamefont {Tianqi}\
  \bibnamefont {Zhao}},\ }\bibfield  {title} {\enquote {\bibinfo {title}
  {Limiting masses and radii of neutron stars and their implications},}\ }\href
  {\doibase 10.1103/PhysRevC.103.045808} {\bibfield  {journal} {\bibinfo
  {journal} {Phys. Rev. C}\ }\textbf {\bibinfo {volume} {103}},\ \bibinfo
  {pages} {045808} (\bibinfo {year} {2021}{\natexlab{b}})},\ \Eprint
  {http://arxiv.org/abs/2009.06441} {arXiv:2009.06441 [nucl-th]} \BibitemShut
  {NoStop}%
\bibitem [{\citenamefont {Contrera}\ \emph {et~al.}(2022)\citenamefont
  {Contrera}, \citenamefont {Blaschke}, \citenamefont {Carlomagno},
  \citenamefont {Grunfeld},\ and\ \citenamefont {Liebing}}]{Contrera:2022tqh}%
  \BibitemOpen
  \bibfield  {author} {\bibinfo {author} {\bibfnamefont {G.~A.}\ \bibnamefont
  {Contrera}}, \bibinfo {author} {\bibfnamefont {D.}~\bibnamefont {Blaschke}},
  \bibinfo {author} {\bibfnamefont {J.~P.}\ \bibnamefont {Carlomagno}},
  \bibinfo {author} {\bibfnamefont {A.~G.}\ \bibnamefont {Grunfeld}}, \ and\
  \bibinfo {author} {\bibfnamefont {S.}~\bibnamefont {Liebing}},\ }\bibfield
  {title} {\enquote {\bibinfo {title} {Quark-nuclear hybrid equation of state
  for neutron stars under modern observational constraints},}\ }\href {\doibase
  10.1103/PhysRevC.105.045808} {\bibfield  {journal} {\bibinfo  {journal}
  {Phys. Rev. C}\ }\textbf {\bibinfo {volume} {105}},\ \bibinfo {pages}
  {045808} (\bibinfo {year} {2022})},\ \Eprint
  {http://arxiv.org/abs/2201.00477} {arXiv:2201.00477 [nucl-th]} \BibitemShut
  {NoStop}%
\bibitem [{\citenamefont {Yang}\ \emph
  {et~al.}(2021{\natexlab{b}})\citenamefont {Yang}, \citenamefont {Pi},
  \citenamefont {Zheng},\ and\ \citenamefont {Weber}}]{Yang:2021sqg}%
  \BibitemOpen
  \bibfield  {author} {\bibinfo {author} {\bibfnamefont {Shu-Hua}\ \bibnamefont
  {Yang}}, \bibinfo {author} {\bibfnamefont {Chun-Mei}\ \bibnamefont {Pi}},
  \bibinfo {author} {\bibfnamefont {Xiao-Ping}\ \bibnamefont {Zheng}}, \ and\
  \bibinfo {author} {\bibfnamefont {Fridolin}\ \bibnamefont {Weber}},\
  }\bibfield  {title} {\enquote {\bibinfo {title} {Constraints from compact
  star observations on non-{{Newtonian}} gravity in strange stars based on a
  density dependent quark mass model},}\ }\href {\doibase
  10.1103/PhysRevD.103.043012} {\bibfield  {journal} {\bibinfo  {journal}
  {Phys. Rev. D}\ }\textbf {\bibinfo {volume} {103}},\ \bibinfo {pages}
  {043012} (\bibinfo {year} {2021}{\natexlab{b}})},\ \Eprint
  {http://arxiv.org/abs/2101.11192} {arXiv:2101.11192 [astro-ph.HE]}
  \BibitemShut {NoStop}%
\bibitem [{\citenamefont {Hinderer}(2008)}]{Hinderer:2007mb}%
  \BibitemOpen
  \bibfield  {author} {\bibinfo {author} {\bibfnamefont {Tanja}\ \bibnamefont
  {Hinderer}},\ }\bibfield  {title} {\enquote {\bibinfo {title} {Tidal {{Love}}
  numbers of neutron stars},}\ }\href {\doibase 10.1086/533487} {\bibfield
  {journal} {\bibinfo  {journal} {Astrophys. J.}\ }\textbf {\bibinfo {volume}
  {677}},\ \bibinfo {pages} {1216--1220} (\bibinfo {year} {2008})},\ \Eprint
  {http://arxiv.org/abs/0711.2420} {arXiv:0711.2420 [astro-ph]} \BibitemShut
  {NoStop}%
\bibitem [{\citenamefont {Flanagan}\ and\ \citenamefont
  {Hinderer}(2008)}]{Flanagan:2007ix}%
  \BibitemOpen
  \bibfield  {author} {\bibinfo {author} {\bibfnamefont {Eanna~E.}\
  \bibnamefont {Flanagan}}\ and\ \bibinfo {author} {\bibfnamefont {Tanja}\
  \bibnamefont {Hinderer}},\ }\bibfield  {title} {\enquote {\bibinfo {title}
  {Constraining neutron star tidal {{Love}} numbers with gravitational wave
  detectors},}\ }\href {\doibase 10.1103/PhysRevD.77.021502} {\bibfield
  {journal} {\bibinfo  {journal} {Phys. Rev. D}\ }\textbf {\bibinfo {volume}
  {77}},\ \bibinfo {pages} {021502} (\bibinfo {year} {2008})},\ \Eprint
  {http://arxiv.org/abs/0709.1915} {arXiv:0709.1915 [astro-ph]} \BibitemShut
  {NoStop}%
\bibitem [{\citenamefont {Fonseca}\ \emph {et~al.}(2021)\citenamefont {Fonseca}
  \emph {et~al.}}]{Fonseca:2021wxt}%
  \BibitemOpen
  \bibfield  {author} {\bibinfo {author} {\bibfnamefont {E.}~\bibnamefont
  {Fonseca}} \emph {et~al.},\ }\bibfield  {title} {\enquote {\bibinfo {title}
  {Refined mass and geometric measurements of the high-mass {{PSR
  J0740}}+6620},}\ }\href {\doibase 10.3847/2041-8213/ac03b8} {\bibfield
  {journal} {\bibinfo  {journal} {Astrophys. J. Lett.}\ }\textbf {\bibinfo
  {volume} {915}},\ \bibinfo {pages} {L12} (\bibinfo {year} {2021})},\ \Eprint
  {http://arxiv.org/abs/2104.00880} {arXiv:2104.00880 [astro-ph.HE]}
  \BibitemShut {NoStop}%
\bibitem [{\citenamefont {Cromartie}\ \emph {et~al.}(2020)\citenamefont
  {Cromartie} \emph {et~al.}}]{NANOGrav:2019jur}%
  \BibitemOpen
  \bibfield  {author} {\bibinfo {author} {\bibfnamefont {H.~T.}\ \bibnamefont
  {Cromartie}} \emph {et~al.} (\bibinfo {collaboration} {NANOGrav}),\
  }\bibfield  {title} {\enquote {\bibinfo {title} {Relativistic {{Shapiro}}
  delay measurements of an extremely massive millisecond pulsar},}\ }\href
  {\doibase 10.1038/s41550-019-0880-2} {\bibfield  {journal} {\bibinfo
  {journal} {Nature Astron.}\ }\textbf {\bibinfo {volume} {4}},\ \bibinfo
  {pages} {72--76} (\bibinfo {year} {2020})},\ \Eprint
  {http://arxiv.org/abs/1904.06759} {arXiv:1904.06759 [astro-ph.HE]}
  \BibitemShut {NoStop}%
\bibitem [{\citenamefont {Holdom}\ \emph {et~al.}(2018)\citenamefont {Holdom},
  \citenamefont {Ren},\ and\ \citenamefont {Zhang}}]{Holdom:2017gdc}%
  \BibitemOpen
  \bibfield  {author} {\bibinfo {author} {\bibfnamefont {Bob}\ \bibnamefont
  {Holdom}}, \bibinfo {author} {\bibfnamefont {Jing}\ \bibnamefont {Ren}}, \
  and\ \bibinfo {author} {\bibfnamefont {Chen}\ \bibnamefont {Zhang}},\
  }\bibfield  {title} {\enquote {\bibinfo {title} {Quark matter may not be
  strange},}\ }\href {\doibase 10.1103/PhysRevLett.120.222001} {\bibfield
  {journal} {\bibinfo  {journal} {Phys. Rev. Lett.}\ }\textbf {\bibinfo
  {volume} {120}},\ \bibinfo {pages} {222001} (\bibinfo {year} {2018})},\
  \Eprint {http://arxiv.org/abs/1707.06610} {arXiv:1707.06610 [hep-ph]}
  \BibitemShut {NoStop}%
\bibitem [{\citenamefont {Son}\ and\ \citenamefont
  {Stephanov}(2001)}]{Son:2000xc}%
  \BibitemOpen
  \bibfield  {author} {\bibinfo {author} {\bibfnamefont {D.~T.}\ \bibnamefont
  {Son}}\ and\ \bibinfo {author} {\bibfnamefont {Misha~A.}\ \bibnamefont
  {Stephanov}},\ }\bibfield  {title} {\enquote {\bibinfo {title} {{{QCD}} at
  finite isospin density},}\ }\href {\doibase 10.1103/PhysRevLett.86.592}
  {\bibfield  {journal} {\bibinfo  {journal} {Phys. Rev. Lett.}\ }\textbf
  {\bibinfo {volume} {86}},\ \bibinfo {pages} {592--595} (\bibinfo {year}
  {2001})},\ \Eprint {http://arxiv.org/abs/hep-ph/0005225}
  {arXiv:hep-ph/0005225} \BibitemShut {NoStop}%
\bibitem [{\citenamefont {Kogut}\ and\ \citenamefont
  {Sinclair}(2002)}]{Kogut:2002zg}%
  \BibitemOpen
  \bibfield  {author} {\bibinfo {author} {\bibfnamefont {J.~B.}\ \bibnamefont
  {Kogut}}\ and\ \bibinfo {author} {\bibfnamefont {D.~K.}\ \bibnamefont
  {Sinclair}},\ }\bibfield  {title} {\enquote {\bibinfo {title} {Lattice
  {{QCD}} at finite isospin density at zero and finite temperature},}\ }\href
  {\doibase 10.1103/PhysRevD.66.034505} {\bibfield  {journal} {\bibinfo
  {journal} {Phys. Rev. D}\ }\textbf {\bibinfo {volume} {66}},\ \bibinfo
  {pages} {034505} (\bibinfo {year} {2002})},\ \Eprint
  {http://arxiv.org/abs/hep-lat/0202028} {arXiv:hep-lat/0202028} \BibitemShut
  {NoStop}%
\end{thebibliography}%

\end{document}